\documentstyle[prd,aps,graphics]{revtex}
\begin{document}
\def\mathrm{ } 
\def\Universita{Universit\'a}
\def\Paris{Par\'\i{}s}
\def\Perez{P\'erez}
\def\Gunther{G\"unther}
\def\Schutzhold{Sch\"utzhold}
\def\Lofstedt{L\"ofstedt}
\def\Garcia{Garc\'\i{}a }
\def\Ruggeberg{R\"uggeberg}
\title{Sonoluminescence as a QED vacuum effect.\\
 II: Finite Volume Effects} 
\author{S. Liberati${}^{\dagger}$}
\address{International School for Advanced Studies, Via Beirut 2-4, 
34014 Trieste, Italy\\
INFN sezione di Trieste}
\author{Matt Visser${}^{\P}$}
\address{Physics Department, Washington University, 
Saint Louis, Missouri 63130-4899, USA}
\author{F. Belgiorno${}^{\ddagger}$}
\address{\Universita\  degli Studi di Milano, Dipartimento di Fisica, 
Via Celoria 16,  20133 Milano, Italy}
\author{D.W. Sciama${}^{\S}$}
\address{International School for Advanced Studies, Via Beirut 2-4, 
34014 Trieste, Italy\\
International Center for Theoretical Physics,  Strada Costiera 11, 
34014 Trieste, Italy\\
Physics Department, Oxford University, Oxford, England\\
\smallskip}

\date{10 May 1999; \LaTeX-ed \today}

\maketitle{}

\begin{abstract}
In a companion paper [quant-ph/9904013] we have investigated several
variations of Schwinger's proposed mechanism for sonoluminescence. We 
demonstrated that any realistic version of Schwinger's mechanism
must depend on extremely rapid (femtosecond) changes in refractive
index, and discussed ways in which this might be physically plausible.
To keep that discussion tractable, the technical computations in
that paper were limited to the case of a homogeneous dielectric
medium. In this paper we investigate the additional complications
introduced by finite-volume effects. The basic physical scenario
remains the same, but we now deal with finite spherical bubbles,
and so must decompose the electromagnetic field into Spherical
Harmonics and Bessel functions. We demonstrate how to set up the
formalism for calculating Bogolubov coefficients in the sudden
approximation, and show that we qualitatively retain the results
previously obtained using the homogeneous-dielectric (infinite
volume) approximation.
\end{abstract}


\section{Introduction}
\def\in{{\mathrm in}}
\def\out{{\mathrm out}}
\def\inside{{\mathrm inside}}
\def\outside{{\mathrm outside}}
\def\liquid{{\mathrm liquid}}
\def\gas{{\mathrm gas}}
\def\ng{n_\gas}
\def\ngi{n_\gas^\in}
\def\ngo{n_\gas^\out}
\def\nl{n_\liquid}
\def\ni{n_\in}
\def\no{n_\out}
\def\nis{n_\inside}
\def\nos{n_\outside}
\def\oi{\omega_\in}
\def\oo{\omega_\out}
\def\Ni{ {\cal N}_\in}
\def\No{ {\cal N}_\out}
\def\max{\hbox{max}}
\def\min{\hbox{min}}
\def\sinc{\mathop{\hbox{sinc}}}
\def\half{{\textstyle{1\over2}}}
\def\quarter{{\textstyle{1\over4}}}

Sonoluminescence (SL)~\cite{Physics-Reports} is a phenomenon whose
underlying physical mechanism is still highly controversial. In a
companion paper~\cite{Companion} we have extensively discussed
Schwinger's proposed mechanism: a mechanism based on changes in the
quantum-electrodynamic (QED) vacuum. We discussed Schwinger's original
version of the model~\cite{Sc1,Sc2,Sc3,Sc4,Sc5,Sc6,Sc7}, the Eberlein
variant~\cite{Eberlein1,Eberlein2,Eberlein3,Eberlein-discussion}, and
proposed a variant of our own~\cite{Companion}. (See also
\cite{Letter,Original,Leipzig,2gamma}.)

One of the key features of photon production by a space-dependent and
time-dependent refractive index is that for a change occurring on a
timescale $\tau$, the amount of photon production is exponentially
suppressed by an amount $\exp(-\omega\tau)$~\cite{Companion}.  The
importance for SL is that the experimental spectrum is {\em not\,}
exponentially suppressed at least out to the far ultraviolet. Thus the
timescale for change in the refractive index must be of order a {\em
femtosecond}, and any Casimir--based model has to take into account
that {\em the change in the refractive index cannot be due just to the
change in the bubble radius}.  To achieve this, we adjust basic
aspects of the model: We will move away from the original Schwinger
suggestion, in that it is no longer the collapse from $R_{\mathrm
max}$ to $R_{\mathrm min}$ that is important. {\em Instead we
postulate a rapid (femtosecond) change in refractive index of the gas
bubble when it hits the van der Waals hard core}~\cite{Companion}.

In this paper we develop the relevant formalism for taking finite
volume effects into account. The quantized electromagnetic field is
decomposed into a set of basis states (spherical harmonics times
radial wavefunctions). The radial wavefunctions are piecewise Bessel
functions with suitable boundary conditions applied at the surface of
the bubble. Using these basis states and the sudden approximation we
can calculate the Bogolubov coefficients relating ``in'' and ``out''
vacuum states. In the infinite volume, the discussion of the companion
paper~\cite{Companion} is recovered, while for physically realistic
finite volumes we see significant but not overwhelming
modifications. The uncertainties in our knowledge of the refractive
index as a function of frequency are also significant.  We find that
we can get both qualitative and approximately quantitative agreement
with the experimentally observed spectrum, but there is a considerable 
amount of unknown condensed-matter physics associated with the details of
the behavior of the refractive index. We feel that the theoretical
calculations have now been pushed as far as is meaningful given the
current state of knowledge, and that further progress will depend on
looking for the experimental signatures discussed in~\cite{Companion}.

\section{Bogolubov coefficients}

To estimate the spectrum and efficiency of photon production we
decided to study a single pulsation of the bubble.  We are not
concerned with the detailed dynamics of the bubble surface.  In
analogy with the subtraction procedure of the static calculations of
Schwinger~\cite{Sc1,Sc2,Sc3,Sc4,Sc5,Sc6,Sc7} or of Carlson {\em et
al.}~\cite{CMMV1,CMMV2,MV} we shall consider two different
configurations.  An ``in'' configuration with a bubble of refractive
index $\ngi$ in a medium of dielectric constant $\epsilon_{\outside}$,
and an ``out'' configuration with a bubble of refractive index $\ngo$
in a medium of dielectric constant $\epsilon_{\outside}$.  These two
configurations will correspond to two different bases for the
quantization of the field.  (For the sake of simplicity we take, as
Schwinger did, only the electric part of QED, reducing the problem to
a scalar electrodynamics).  The two bases will be related by Bogolubov
coefficients in the usual way.  Once we determine these coefficients
we easily get the number of created particles per mode, and from this
the spectrum.  (This calculation uses the ``sudden
approximation'': Changes in the refractive index are assumed to be
non-adiabatic, see references
\cite{Companion,Letter,Original,Leipzig,2gamma} for more discussion.)

Let us adopt the Schwinger formalism and consider the equations of the
electric field in spherical coordinates and with a time-independent
dielectric constant. (We temporarily set $c=1$ for ease of notation,
and shall reintroduce appropriate factors of the speed of light when
needed for clarity.)  Then in the asymptotic future and asymptotic
past, where the refractive index is taken to be time-independent, we
are interested in solving
\begin{equation}
\label{E:static}
\epsilon(r) \; \partial_{0}(\partial_{0} E)-\nabla^{2} E=0,
\end{equation}
with $\epsilon(r)$ being piecewise constant.  We look for
solutions of the form
\begin{equation}
E=\Phi(r,t) \; Y_{lm}(\Omega) \; {1\over r}.
\end{equation}
Then one finds
\begin{equation}
\epsilon(\partial^{2}_{0}\Phi)-(\partial_{r}^{2} \Phi)+{1\over r^{2}}
l(l+1) \Phi =0.
\label{E:eqm2}
\end{equation}
For both the ``in'' and ``out'' solution the field equation in $r$
is given by:
\begin{equation}
\epsilon\partial_{0}^{2} \Phi-\partial_{r}^{2}\Phi +{1\over r^{2}} l(l+1)
\Phi=0.
\end{equation}
In both asymptotic regimes (past and future) one has a static
situation (a bubble of dielectric $\ngi$ in the dielectric
$\nl$, or a bubble of dielectric $\ngo$ in the dielectric $\nl$)
so one can in this limit factorize the time and radius dependence
of the modes:  $\Phi(r,t)=e^{i\omega t} f(r)$.  One gets
\begin{equation}
f^{''}+\left (\epsilon \omega^{2} -{1\over r^{2}} l(l+1)\right) f=0.
\end{equation}
This is a well known differential equation. To handle it more easily
in a standard way we can cast it as an eigenvalues problem
\begin{equation}
f^{''}-\left( {1\over r^{2}} l(l+1) \right)f=-\kappa^{2}f,
\end{equation}
where $\kappa^2=\epsilon \omega^{2}$.  With the change of variables
$f=r^{1/2} G$, so that $\Phi(r,t)=e^{i\omega t} r^{1/2} G(r)$, we get
\begin{equation}
G^{''}+{1\over r}G^{'}+\left(\kappa^{2}-{\nu^2 \over r^{2}} \right)G=0.
\end{equation}
This is the standard Bessel equation. It admits as solutions the
Bessel and Neumann functions of the first type, $J_{\nu}(\kappa
r)$ and $N_{\nu}(\kappa r)$, with $\nu=l+1/2$.  Remember that for
those solutions which have to be well-defined at the origin, $r=0$,
regularity implies the absence of the Neumann functions.  For both the
``in'' and the ``out'' basis we have to take into account that the
dielectric constant changes at the bubble radius ($R$). In fact we have
\begin{equation}
\epsilon^\in=\left\{ 
\begin{array}{llllll}
\vphantom{\Bigg|}
\epsilon_\inside^\in 
& = & (\ngi)^2
& = & 
\mbox{dielectric constant of air-gas mixture} & 
\mbox{if $r\leq R$},\\
\vphantom{\Bigg|}
\epsilon_\outside^\in 
& = & n^2_\liquid 
& = & 
\mbox{dielectric constant of ambient liquid (typically water)} & 
\mbox{if $r > R$}.
\end{array}
\right.
\end{equation}
After the change in refractive index, we have
\begin{equation}
\epsilon^\out=\left\{ 
\begin{array}{llllll}
\vphantom{\Bigg|}
\epsilon_\inside^\out 
& = & (\ngo)^2
& = & 
\mbox{dielectric constant of air-gas mixture} & 
\mbox{if $r\leq R$},\\
\vphantom{\Bigg|}
\epsilon_\outside 
& = & n^2_\liquid 
& = & 
\mbox{dielectric constant of ambient liquid (typically water)} & 
\mbox{if $r > R$}.
\end{array}
\right.
\end{equation}

In the original version of the Schwinger model it was usual to
simplify calculations by using the fact that the dielectric constant
of air is approximately equal 1 at standard temperature and pressure
(STP), and then dealing only with the dielectric constant of water
($n_\liquid = \sqrt{\epsilon_\outside} \approx 1.3$). We wish to avoid
this temptation on the grounds that the sonoluminescent flash is known
to occur within $500$ picoseconds of the bubble achieving minimum
radius. Under these conditions the gases trapped in the bubble are
close to the absolute maximum density implied by the hard core
repulsion incorporated into the van der Waals equation of state.  Gas
densities are approximately one million times atmospheric and
conditions are nowhere near STP.  (For details, see~\cite{Holzfuss}
page 5437, and the discussion in~\cite{Companion}.) For this reason
we shall explicitly keep track of $n_\gas^\in$, $n_\gas^\out$, and
$n_\liquid$ in the formalism we develop.  Defining the ``in'' and
``out'' frequencies, $\omega_\in$ and $\omega_\out$ respectively, one
has
\begin{equation}
G^\in_{\nu}(\ngi,\nl,\omega_\in,r)
=\left \{
\begin{array}{ll}
\vphantom{\Bigg|}
\Xi_\nu^\in \; A_{\nu}^\in \; 
J_{\nu}(\ngi\, \omega_\in r) & 
\mbox{if $r\leq R$},\\ 
\vphantom{\Bigg|}
\Xi_\nu^\in \; \left[
B_{\nu}^\in \; J_{\nu}(\nl\, \omega_\in r)+
C_{\nu}^\in \; N_{\nu}(\nl\, \omega_\in r)
\right]&
\mbox{if $r > R$.}
\end{array}
\right.
\end{equation}
Here $\Xi_\nu^\in$ is an overall normalization. The $A_\nu^\in$,
$B_\nu^\in$, and $C_\nu^\in$ coefficients are determined by the
matching conditions at $R$
\begin{equation}
\begin{array}{lll}
\vphantom{\Bigg|}
A_{\nu}^\in \; J_{\nu}(\ngi\, \omega_\in  R)&=&
B_{\nu}^\in \; J_{\nu}(\nl\, \omega_\in R)+ 
C_{\nu}^\in \; N_{\nu}(\nl\, \omega_\in R),
\\
\vphantom{\Bigg|}
A_{\nu}^\in \; J_{\nu}{'}(\ngi\, \omega_\in R)&=&
B_{\nu}^\in \; J_{\nu}{'}(\nl\, \omega_\in R)+
C_{\nu}^\in \; N_{\nu}{'}(\nl\, \omega_\in R),
\end{array}
\label{E:coef}
\end{equation}
(the primes above denote derivatives with respect to $r$),
together with the convention that
\begin{equation}
|B|^2 + |C|^2 = 1.
\end{equation}
The ``out'' basis is easily obtained solving the same equations but
systematically replacing $\ngi$ by $\ngo$. There will be additional
coefficients, $\Xi_\nu^\out$, $A_\nu^\out$, $B_\nu^\out$, and
$C_\nu^\out$, corresponding to the ``out'' basis.

(In an earlier work~\cite{Original}
[quant-ph/9805031] we adopted a normalization convention such that
$\Xi_\nu^\out$ was always equal to unity. Though that convention is
physically equivalent to this one (there are compensating factors in
the phase space measure), in this paper we find it more convenient to
explicitly keep track of this overall prefactor because it helps us to
compare our finite volume results to the analytically tractable
homogeneous model discussed in~\cite{Companion}.)

We now demand the existence of a normalized scalar product such
that we can define orthonormal eigenfunctions
\begin{equation}
(\Phi^{i},\Phi^{j})=\delta^{ij}.
\end{equation}
For the special case $\ng=\nl$, (corresponding to a completely
homogeneous space, in which case $A=1=B$, $C=0$), it is useful to
adopt a slight variant of the ordinary scalar product, defined by
\begin{equation} 
(\phi_{1},\phi_{2})=
i \, n^2 \int_{\Sigma_t}
\phi_{1}^* \stackrel{\leftrightarrow}{\partial}_{0}\phi_{2}
\: d^{3}x,
\end{equation}
though we shall soon see that this definition will need to be
generalized for a time-dependent and position-dependent refractive
index. (See Appendix A, and compare to the discussion
in~\cite{Companion}). If we now take the scalar product of two
eigenfunctions, we expect to obtain a normalization condition which
can be written as
\begin{equation}
\left(\Phi^{i}_{[\ng=\nl]},\Phi^{j}_{[\ng=\nl]}\right) = \delta^{ij}.
\end{equation}
Inserting the explicit form of the $\Phi$ functions 
we get 
\begin{eqnarray}
\left(\Phi^{i}_{[\ng=\nl]},\Phi^{j}_{[\ng=\nl]}\right)
&=&
\Xi_i^* \; \Xi_j \; \delta_{ll'} \; \delta_{mm'} \; 
n^2 (\omega_i+\omega_j) 
\int_{0}^{\infty} r dr \;
J_{\nu}(n \omega_i r) \; J_{\nu}(n \omega_j r) \;
e^{i(\omega_i-\omega_j)t}\\
&=& 
2 \; \Xi_i^* \; \Xi_j \; \delta_{ll'} \; \delta_{mm'} \; 
n^2 \; (\omega_i+\omega_j) \;
\frac{\delta(n \omega_i- n\omega_j)}{n \omega_i + n\omega_j} \;
e^{i(\omega_i-\omega_j)t}
\\
&=& 
2 \; \Xi_i^* \; \Xi_j \; \delta_{ll'} \; \delta_{mm'} \; 
n \; \delta(\kappa_i - \kappa_j),
\end{eqnarray} 
where we have used the inversion formula for Hankel Integral
transforms~\cite{Bateman,Jackson}, which can be written as
\begin{equation}
\label{E:hankel}
\int_{0}^{\infty}   r dr \;
J_{\nu}(\kappa_1 r) \; J_{\nu}(\kappa_2 r) =
{\delta(\kappa_1 - \kappa_2) \over\sqrt{\kappa_1\;\kappa_2}}=
2 \; {\delta(\kappa_1-\kappa_2)\over(\kappa_1+\kappa_2)} =
2 \;  \delta(\kappa_1^2 - \kappa_2^2),
\end{equation}
this result being valid for $Re(\nu) > - \half$, and 
$\kappa_{(1,2)} > 0$.

We now compare this to the behaviour of the three-dimensional delta
function in momentum space
\begin{eqnarray}
\delta^3 (\vec \kappa_i - \vec \kappa_j ) 
&=&
{\delta (\kappa_i - \kappa_j ) \over \kappa_i \; \kappa_j} \; 
\delta^2(\hat \kappa_i - \hat \kappa_j) 
\\
&=&
{\delta (\kappa_i - \kappa_j ) \over \kappa_i \; \kappa_j} \; 
\sum_{lm} Y^*_{lm} (\theta_i,\phi_i) \; Y_{lm}(\theta_j,\phi_j)
\\
&\to&
{\delta (\kappa_i - \kappa_j ) \over \kappa_i \; \kappa_j} \; \delta_{ll'} \; \delta_{mm'},
\end{eqnarray}
to deduce that for homogeneous spaces the most useful normalization is
\begin{equation}
2 \; \Xi_i^* \; \Xi_j \; n  = {1\over \kappa_i \; \kappa_j}.
\end{equation}
This strongly suggests that even for static but non-homogeneous
dielectric configurations it will be advantageous to set
\begin{equation}
\left|\,\Xi^i\right| = {1\over\sqrt{2 n} \; \kappa_i}.
\end{equation}
where $n$ is now the refractive index at spatial infinity.

In our calculations the phase of $\Xi$ is never physically
important. If desired it can be fixed by using the well-known
decomposition of the plane-waves into spherical harmonics and Bessel
functions. See, {\em e.g.}  Jackson~\cite{Jackson} pages 767 and 740,
equations (16.127) and (16.9). The properly normalized plane-wave
states of the companion paper~\cite{Companion} are
\begin{equation}
E(\vec x,t) = {1\over(2\pi)^{3/2}} \; 
{\exp(i[\vec \kappa\cdot \vec x - \omega t])\over\sqrt{2 \; \omega}\; n}
\end{equation}
Using the spherical decomposition of plane waves this equals
\begin{equation}
E(\vec x,t) = 
\sum_{l=0}^\infty {i^l \over \sqrt{2\; n} \; \kappa} \; 
{J_\nu(\kappa r)\over\sqrt{r}} \; \exp(-i\omega t) \;
\sum_{m=-l}^{m=+l} 
Y_{lm}^*(\theta(\hat x),\phi(\hat x)) \; 
Y_{lm}  (\theta(\hat \kappa),\phi(\hat \kappa)).
\end{equation}
This allows us to identify
\begin{equation}
\Xi_l = {i^l \over \sqrt{2\; n} \; \kappa},
\end{equation}
thereby fixing the phase, and verifying our normalization from another
point of view.

To confirm that this is still the most appropriate normalization for
non-homogeneous dielectrics requires a brief digression: The fact that
even for homogeneous spaces the usual inner product needs an explicit
factor of $n^{2}$ to get the correct normalization is our first signal
that the inner product should be somewhat modified for
position-dependent and time-dependent refractive indices. If we now
consider the case of a time-independent but position-dependent
refractive index, then as explained in Appendix A, the inner product
must be generalized to
\begin{equation} 
(\phi_{1},\phi_{2})=
i\, \int_{\Sigma_t} \epsilon(r) \; \phi_{1}^*
\stackrel{\leftrightarrow}{\partial}_{0}\phi_{2} \: d^{3}x,
\end{equation}
Taking the scalar product of a pair of eigenfunctions, inserting the
explicit form of the $\Phi$ functions, and recalling that $\kappa = n
\omega$, we get
\begin{eqnarray}
\label{E:normalization}
(\Phi^{i},\Phi^{j})&=&
(\omega_i+\omega_j) e^{i(\omega_i-\omega_j)t}
\Bigg[ 
\int_{0}^{R} r dr \; (\ng)^2 \; \Xi_i^* \; A_i^* \; \Xi_j \; A_j \;
J_{\nu}(\kappa_i^\gas r)J_{\nu}(\kappa_j^\gas r) 
\nonumber\\
&&\qquad
+
\int_{R}^{\infty} r dr \; (\nl)^2 \; 
\Xi_i^* \;
\left[B_i^* J_\nu(\kappa_i^\liquid r) + 
      C_i^* N_\nu(\kappa_i^\liquid r)\right] 
\nonumber\\
&&\qquad\qquad\qquad
\Xi_j \;
\left[B_j J_\nu(\kappa_j^\liquid r) + 
      C_j N_\nu(\kappa_j^\liquid r)\right] 
\Bigg]
\\
&=& 
2 \; (\omega_i+\omega_j) \;
\Xi_i^* \; \Xi_j \; \{ B_i^* B_j + C_i^* C_j \} \; 
(\nl)^2 \;
\frac{\delta(\kappa_i^\liquid-\kappa_j^\liquid)}
{\kappa_i^\liquid+\kappa_j^\liquid}  \;
e^{i(\omega_i-\omega_j)t}\\
&=& 
2 \; (\omega_i+\omega_j) \;  
\Xi_i^* \; \Xi_j \; \{ B_i^* B_j + C_i^* C_j \} 
\frac{\delta(\omega_i-\omega_j)}{\omega_i+\omega_j} \;
e^{i(\omega_i-\omega_j)t}\\
&=& 
2 \;  \Xi_i^* \; \Xi_j \; 
\{ B_i^* B_j + C_i^* C_j \} \; 
\delta(\omega_i-\omega_j).
\end{eqnarray} 
The last few lines follow from some Bessel function identities we have
collected into Appendix B. These identities can be derived as
generalizations of the Hankel integral transform formula
(\ref{E:hankel}), or via use of specific spectral representations of
the delta function. There are delicate cancellations between surface
terms at $R^+$ and $R^-$, and the subtle part of the calculation
involves the surface term at spatial infinity. This calculation is
most useful in that it verifies for us the normalization condition we
need on the ``in'' and ``out'' asymptotic states: If we adopt the
convention that
\begin{equation}
|B|^2 + |C|^2 = 1,
\end{equation} 
then proper normalization of the wavefunctions demands
\begin{equation}
\left|\Xi^i\right| = {1\over\sqrt{2\,\nl} \; \kappa_i},
\end{equation}
and we see that it is indeed the refractive index at spatial infinity
that is the relevant one for this overall normalization.

We are now ready to ask what happens if we change the refractive index
by making $\epsilon(r,t)$ a function of both position and time. We are
interested in solving the equation
\begin{equation}
\label{E:change}
\partial_{0}(\epsilon(r,t) \; \partial_{0} E)-\nabla^{2} E=0,
\end{equation}
which is the relevant generalization of (\ref{E:static}) to the
time-dependent case.  It should be no surprise that this change affects
both the conserved density and flux, and that the inner product must
again be modified (in what is now a rather obvious fashion). See
Appendix A for the details.
\begin{equation} 
(\phi_{1},\phi_{2}) =
i\, \int_{\Sigma_t} \epsilon(r,t) \; \phi_{1}^*
\stackrel{\leftrightarrow}{\partial}_{0}\phi_{2}\: d^{3}x,
\end{equation}
The Bogolubov coefficients (relative to this inner product) can
now be {\em defined} as
\begin{eqnarray}
\alpha_{ij}
&=&
({E_{i}^\out},{E_{j}^\in}),
\\
\beta_{ij}
&=&(
{E_{i}^\out}^*, {E_{j}^\in}).
\end{eqnarray}
Where $E_{j}^\in$ now denotes an {\em exact} solution of
the time-dependent equation (\ref{E:change}) that in the infinite past
approaches a solution of the static equation (\ref{E:static}) with
$\epsilon \to \epsilon_\in(r)$ and eigen-frequency
$\omega_j$. Similarly $E_{i}^\out$ now denotes an {\em exact}
solution of the time-dependent equation (\ref{E:change}) that in the
infinite future approaches a solution of the static equation
(\ref{E:static}) with $\epsilon \to \epsilon_\out(r)$ and
eigen-frequency $\omega_i$. The inner product used to define the
Bogolubov coefficients has been carefully arranged to correspond to a
``conserved charge''. With the conventions we have in place the
absolute values of the Bogolubov coefficient are {\em independent} of
the choice of time-slice $\Sigma_t$ on which the spatial integral is
evaluated. With minor modifications, as explained in Appendix A, the
inner product can further be generalized to enable it to be defined for
any arbitrary edgeless achronal spacelike hypersurface, not just the
constant time time-slices.  (This whole formalism is very closely
related to the S-matrix formalism of quantum field theories, where the
S-matrix relates asymptotic ``in'' and ``out'' states.)

Of course, evaluating the Bogolubov coefficients involves solving
the {\em exact} time-dependent problem (\ref{E:change}), subject to
the specified boundary conditions, a task that is in general
hopeless. It is at this stage that we shall explicitly invoke the
sudden approximation by choosing the dielectric constant to
be\footnote{
This was implicit in the earlier work~\cite{Original}
[quant-ph/9805031]. We make it explicit here since this point has
caused some confusion.}
%
\begin{equation}
\label{E:eps-t}
\epsilon(r,t) = 
\epsilon_\in(r) \; \Theta(-t) + 
\epsilon_\out(r) \; \Theta(t).
\end{equation}
This is a simple step-function transition from $\epsilon_\in(r)$ to
$\epsilon_\out(r)$ at time $t=0$. For $t<0$ the exact eigenstates are
given in terms of the static problem with $\epsilon= \epsilon_\in(r)$,
and for $t>0$ the exact eigenstates are given in terms of the static
problem with $\epsilon=\epsilon_\out(r)$. To evaluate the Bogolubov
coefficients in the simplest manner, we chose the spacelike
hypersurface to be the $t=0$ hyperplane. The inner product then
reduces to
\begin{equation} 
(\phi_{1},\phi_{2}) =
i\; \int_{t=0} 
\epsilon(r,t=0) \;
\phi_{1}^{*} 
\stackrel{\leftrightarrow}{\partial}_{0}
\phi_{2} \: d^{3}x,
\end{equation}
with the relevant eigenmodes being those of the {\em static} ``in''
and ``out'' problems. (At a fundamental level, this formalism is
just a slight modification of the standard machinery of the sudden
approximation in quantum mechanical perturbation theory.)\\ 
There is actually a serious ambiguity hiding here: What value are we to
assign to $\epsilon(r,t=0)$? One particularly simple candidate is
\begin{equation}
\epsilon(r,t=0) 
\to 
\half 
\left[ \epsilon_\in(r) + \epsilon_\out(r) \right]
=
\half 
\left[ n_\in(r)^2 + n_\out(r)^2 \right],
\end{equation}
but this candidate is far from unique. For instance, we could
rewrite (\ref{E:eps-t}) as
\begin{equation}
\epsilon(r,t) = 
\exp\Big( 
\ln\{\epsilon_\in(r)\}  \; \Theta(-t) + 
\ln\{\epsilon_\out(r)\} \; \Theta(t)
\Big).
\end{equation}
For $t\neq0$ this is identical to (\ref{E:eps-t}), but for
$t=0$ this would more naturally lead to the prescription
\begin{equation}
\epsilon(r,t=0) 
\to 
\sqrt{\epsilon_\in(r) \epsilon_\out(r)}
= 
n_\in(r) n_\out(r).
\end{equation}
By making a comparison with the analytic calculation for homogeneous
media presented in~\cite{Companion} we shall in fact show that this is
the correct prescription, but for the meantime will simply adopt the
notation
\begin{equation}
\epsilon(r,t=0) =
\gamma\left(n_\in(r);n_\out(r)\right),
\end{equation}
where the only property of $\gamma(n_1;n_2)$ that we really need
to use at this stage is that when $n_1 =  n_2$
\begin{equation}
\gamma(n;n) = n^2.
\end{equation}
(This property follows automatically from considering the
static time-independent case.)

We are mainly interested in the Bogolubov coefficient $\beta$, since
it is $|\beta|^{2}$ that is linked to the total number of particles
created.  By a direct substitution it is easy to find the expression:
\begin{eqnarray}
\beta_{ll',mm'}(\omega_\in,\omega_\out) 
&=&
i \int_{0}^{\infty} 
\gamma\left(\ni(r),\no(r)\right)
\left( \Phi_\out(r,t) \; Y_{lm}(\Omega)  \; {1\over r} \right) 
\stackrel{\leftrightarrow}{\partial}_{0}
\left( \Phi_\in(r,t) \; Y_{l^{\prime}m^{\prime}}(\Omega)  
\;{1\over r}
\right) \: r^2 dr d\Omega,\\
&=& -(\omega_\in-\omega_\out) \; 
e^{i(\omega_\out+\omega_\in)t} 
\delta_{l l^{\prime}}\; \delta_{m,-m^{\prime}} \; 
\nonumber\\
&& 
\qquad \qquad \times
\int_{0}^{\infty}
\gamma\left(\ni(r);\no(r)\right) \;
G^\out_{l}(\ngo,\nl,\omega_\out,r) \; 
G^\in_{l^{\prime}}(\ngi,\nl,\omega_\in,r) \: r dr.
\label{E:beta1}
\end{eqnarray}
(The $\delta_{m,-m'}$ arises because of the {\em absence } of
a relative complex conjugation in the angular integrals for
$\beta$. On the other hand, the $\alpha$ coefficient will be
proportional to $\delta_{mm'}$.) To compute the radial
integral one needs some ingenuity, let us write the equations of
motion for two different values of the eigenvalues, $\kappa_1$ and
$\kappa_2$.
\begin{eqnarray}
G_{\kappa_1}^{''}+{1\over r}G_{\kappa_1}^{'}
+\left (\kappa_1^{2}-
{1\over r^{2}} (l+\half)^2  \right)G_{\kappa_1}
&=&0,
\\
G_{\kappa_2}^{''}+{1\over r}G_{\kappa_2}^{'}
+\left (\kappa_2^{2}-
{1\over r^{2}} (l+\half)^2 \right)G_{\kappa_2}&=&0.
\end{eqnarray}
If we multiply the first by $G_{\kappa_2}$ and the second by
$G_{\kappa_1}$ we get
\begin{eqnarray}
G_{\kappa_1}^{''}G_{\kappa_2}+{1\over r}G_{\kappa_1}^{'}G_{\kappa_2}
+ \left  (\kappa_1^{2}-
{1\over r^{2}} (l+\half)^2\right)G_{\kappa_1}G_{\kappa_2}
&=&0,
\\
G_{\kappa_2}^{''}G_{\kappa_1}+{1\over r}G_{\kappa_2}^{'}G_{\kappa_1}
+ \left (\kappa_2^{2}-
{1\over r^{2}} (l+\half)^2 \right)G_{\kappa_1}G_{\kappa_1}&=&0.
\end{eqnarray}
Subtracting the second from the first we then obtain
\begin{equation}
\left( G_{\kappa_1}^{''}G_{\kappa_2}-G_{\kappa_2}^{''}G_{\kappa_1}\right)+
{1\over r}\left(
G_{\kappa_1}^{'}G_{\kappa_2}-G_{\kappa_2}^{'}G_{\kappa_1}
\right)+
(\kappa_2^{2}-\kappa_1^{2}) G_{\kappa_1}G_{\kappa_2}=0.
\end{equation}
The second term on the left hand side is a pseudo--Wronskian determinant
\begin{equation}
W_{\kappa_1\kappa_2}(r)
=
G_{\kappa_1}^{'}(r) G_{\kappa_2}(r) - G_{\kappa_2}^{'}(r) G_{\kappa_1}(r),
\end{equation}
and the first term is its total derivative $dW_{\kappa_1\kappa_2}/dr$.
(This is a pseudo-Wronskian, not a true Wronskian, since the two
functions $G_{\kappa_1}$ and $G_{\kappa_2}$ correspond to different
eigenvalues and so solve different differential equations.) The
derivatives are all with respect to the variable $r$.  Using this
definition we can cast the integral over $r$ of the product of two
given solutions into a simple form. Generically:
\begin{equation}
\left(\kappa_2^{2}-\kappa_1^{2}\right) \int_{a}^{b} r dr\; G_{\kappa_1}G_{\kappa_2}
=
\int_{a}^{b} rdr \; \frac{dW_{\kappa_1\kappa_2}}{dr}+ \int_{a}^{b} 
dr\; W_{\kappa_1\kappa_2}.
\end{equation}
That is
\begin{equation}
\int_{a}^{b} r dr\; G_{\kappa_1}G_{\kappa_2}
=
{1\over
{\kappa_2^{2}-\kappa_1^{2}}}
\left [ 
\left. W_{\kappa_1\kappa_2}\:r \right|_{a}^{b} 
-\int_{a}^{b} dr \;
W_{\kappa_1\kappa_2}+ \int_{a}^{b} dr \; W_{\kappa_1\kappa_2}.
\right]
\end{equation}
So the final result is
\begin{equation}
\int_{a}^{b} r dr \; G_{\kappa_1}G_{\kappa_2}= 
\left.
{1\over {\kappa_2^{2}-\kappa_1^{2}} } \; 
\left( W_{\kappa_1\kappa_2} \; r\right) 
\right|^{b}_{a}.
\end{equation}

This expression can be applied (piecewise) in our specific case
[equation (\ref{E:beta1})]. We obtain:
\begin{eqnarray}
\int^{\infty}_{0} &r&  dr \: 
\gamma\left(\ni(r);\no(r)\right) \; 
G^\out_{\nu}(\ngo,\nl,\omega^\out,r) \; 
G^\in_{\nu}(\ngi,\nl,\omega^\in,r)
\\
&=&\int^{R}_{0} r\; dr\: 
\gamma\left(\ngi;\ngo\right) \;
G^\out_{\nu}(\ngo\,\omega_\out r)
G^\in_{\nu}(\ngi\,\omega_\in r)
\nonumber\\
&&\qquad
+\int_{R}^{\infty} r \; dr \: (\nl)^2 \; 
G^\out_{\nu}(\nl\,\omega_\out r) 
G^\in_{\nu}(\nl\,\omega_\in r)
\\
&=& 
\gamma\left(\ngi;\ngo\right) \; 
\frac{
\left\{ r 
W[G^\out_{\nu}(\ngo\,\omega_\out r),
G^\in_{\nu}(\ngi\,\omega_\in r)]
\right\}^{R}_{0}
}
{(\ngo\,\omega_\out)^2-(\ngi\,\omega_\in)^2} 
\nonumber\\
&& 
\qquad
+
(\nl)^2
\frac{
\left\{r 
W[G^\out_{\nu}(\nl\,\omega_\out r),
G^\in_{\nu}(\nl\,\omega_\in r)]
\right\}^{\infty}_{0}
}
{(\nl\,\omega_\out)^2-(\nl\,\omega_\in)^2}\\
&=& R\,
\Bigg[
\gamma\left(\ngi;\ngo\right)
\frac{
W[G^\out_{\nu}(\ngo\,\omega_\out r),
G^\in_{\nu}(\ngi\,\omega_\in r)]_{R_{-}}
}
{(\ngo\,\omega_\out)^2-(\ngi\,\omega_\in)^2}
\nonumber\\
&&
\qquad 
-
\frac{
W[G^\out_{\nu}(\nl\,\omega_\out r),
G^\in_{\nu}(\nl\,\omega_\in r)]_{R_{+}}
}
{(\omega_\out)^2-(\omega_\in)^2}
\Bigg],
\end{eqnarray}
where we have used the fact that the above forms are well behaved
(and equal to $0$) for $r=0$. There is an additional delta-function
contribution, proportional to $\delta(\omega_\in -
\omega_\out)$, arising from spatial infinity $r=\infty$.
In the case of the $\beta$ Bogolubov coefficient this can quietly
be discarded because of the explicit $ (\omega_\in -
\omega_\out)$ prefactor\footnote{%
We wish to thank Joshua Feinberg for some insightful questions on this
point that caused us to delve into this issue more deeply.}.
For the $\alpha$ Bogolubov coefficient we would need to explicitly
keep track of this delta-function contribution, since it is ultimately
responsible for the correct normalization of the eigenmodes if we were
to take $\ngo \to \ngi$.  (Here and henceforth we shall automatically
give the same $l$ value to the ``in'' and ``out'' solutions by using
the fact that equation (\ref{E:beta1}) contains a Kronecker delta in
$l$ and $l^{\prime}$.) Finally the two pseudo-Wronskians above are
actually equal (by the junction condition (\ref{E:coef})).  This
equality allows to rewrite integral in equation (\ref{E:beta1}) in a
more compact form
\begin{eqnarray}
\int^{\infty}_{0} &r& \; dr \:
\gamma\left(\ni(r);\no(r)\right) \; 
G^\out_{\nu}(\ngo,\nl,\omega_\out,r) \; 
G^\in_{\nu}(\ngi,\nl,\omega_\in,r)
\nonumber\\
&=& 
\Xi_\in \; \Xi_\out \; A_{\nu}^\in \; A_{\nu}^\out R \;
\left[ 
\frac{\gamma\left(\ngi;\ngo\right)}
{(\ngo\,\omega_\out)^2-(\ngi\,\omega_\in)^2}-
\frac{1}{(\omega_\out)^2-(\omega_\in)^2}
\right]
\nonumber\\
&&
\qquad\qquad
\times
W[J_{\nu}(\ngo\,\omega_\out r),
  J_{\nu}(\ngi\,\omega_\in r)]_{R}\\
&=&
\Xi_\in \; \Xi_\out \;
A_{\nu}^\in \; A_{\nu}^\out R \;
{
[
\{\gamma\left(\ngi;\ngo\right) - (\ngo)^2 \}
\omega_\out^2 
-
\{\gamma\left(\ngi;\ngo\right) - (\ngi)^2 \}
\omega_\in^2
]
\over 
[\omega_\out^2-\omega_\in^2]
}\:
\nonumber\\
&&
\qquad\qquad
\times \frac{
W[J_{\nu}(\ngo\,\omega_\out r),
J_{\nu}(\ngi\,\omega_\in r)]_{R}
}
{[(\ngo\,\omega_\out)^2-(\ngi\,\omega_\in)^2]}.
\end{eqnarray}
Inserting this expression into equation  (\ref{E:beta1}) we get
\begin{eqnarray}
\beta_{lm,l'm'}(\omega_\in,\omega_\out)&=&
\Xi_\in \; \Xi_\out \;
A_\nu^\in \; A_\nu^\out \; R\:
\delta_{l l^{\prime}} \;
\delta_{m,-m^{\prime}} \;
\frac{ 
[
\{\gamma\left(\ngi;\ngo\right) - (\ngo)^2 \}
\omega_\out^2 
-
\{\gamma\left(\ngi;\ngo\right) - (\ngi)^2 \}
\omega_\in^2
]
}
{\omega_\out+\omega_\in} \;
\nonumber\\
&&
\qquad \times
\frac{
W[J_{\nu}(\ngo\,\omega_\out r),
J_{\nu}(\ngi\,\omega_\in r)]_{R}} 
{[(\ngo\,\omega_\out)^2-(\ngi\,\omega_\in)^2]}
\; 
e^{i(\omega_\out+\omega_\in)t}.  
\end{eqnarray}
As a consistency check, this expression has the desirable property
that $\beta\to0$ as $\ngo\to\ngi$: That is, if there is no change in
the refractive index, there is no particle production.  We are mainly
interested in the square of this coefficient summed over $l$ and
$m$. It is in fact this quantity that is linked to the spectrum of the
``out'' particles present in the ``in'' vacuum, and it is this
quantity that is related to the total energy emitted.  Including all
appropriate dimensional factors ($c$, $\hbar$) we would have (in a
plane wave basis)
\begin{equation}
{dN(\vec \kappa_\out^{\;\liquid})\over d^3 \vec \kappa_\out^{\; 
\liquid}} 
=\int|\beta(\vec \kappa_\in^{\; \liquid},\vec
\kappa_\out^{\; \liquid})|^2
\; d^3 \vec \kappa_\in^{\; \liquid}.
\end{equation}
Here, since are are interested in the asymptotic behaviour of the
photons after they escape from the bubble and move to spatial
infinity, we have been careful to express the wave-vectors in terms of
the refractive index of the ambient liquid. This is equivalent
to\footnote{%
Remember that when the photons cross the gas-liquid interface their
frequency, though not their wave-number, is conserved. So we do not
need to distinguish $\omega_\gas$ from $\omega_\liquid$.}
%
\begin{equation}
{dN(\vec \kappa_\out^{\; \liquid})\over d \kappa_\out^{\liquid}} 
= \int \left|\beta(\vec \kappa_\in^{\; \liquid},\vec
\kappa_\out^{\;\liquid})\right|^2
\; \left(\kappa_\in^\liquid\right)^2 \; \left(\kappa_\out^\liquid
\right)^2 \; 
d \kappa_\in^\liquid \; d^2\Omega_\in \; d^2\Omega_\out.
\end{equation}
If we now convert this to a spherical harmonic basis the angular
integrals must be replaced by sums over $l,l'$ and $m,m'$. Furthermore
we can also replace the $d \kappa_\in$ and $d \kappa_\out$ by the associated
frequencies $d\omega_\in$ and $d\omega_\out$ to obtain
\begin{equation}
{dN(\omega_\out)\over d\omega_\out} 
=\int \sum_{ll'} \sum_{mm'} 
\left|\beta_{ll',mm'}(\omega_\in,\omega_\out)\right|^2 \; \nl \;
\nl
\; \left(\kappa_\in^\liquid \right)^2 \; \left(\kappa_\out^\liquid
\right)^2 \; d \omega_\in \;.
\end{equation}
In view of our previous definition of the $\Xi$ factors this implies
\begin{equation}
\label{E:N-spectrum}
{dN(\omega_\out)\over d\omega_\out} 
= \quarter \int 
{|\beta(\omega_\in,\omega_\out)|^2 \over |\Xi_\in|^2 \; |\Xi_\out|^2 } \;
d \omega_\in,
\end{equation}
where we have now defined
\begin{equation}
\left|\beta(\omega_\in,\omega_\out)\right|^2
= \sum_{lm}\sum_{l^{\prime}m^{\prime}}
\left[
\beta_{lm,l^{\prime}m^{\prime}}(\omega_\in,\omega_\out)
\right]^2.
\end{equation}
Note that the normalization factors $\Xi$ quietly cancel out of the
physically observable number spectrum. Other quantities of physical
interest are
\begin{equation}
\label{E:N}
N
= \int {dN(\omega_\out)\over d\omega_\out} \; d\omega_\out,
\end{equation}
and
\begin{equation}
\label{E:E}
E= \hbar \int {dN(\omega_\out)\over d\omega_\out} 
\; \omega_\out \; d\omega_\out.
\end{equation}
Hence we shall concentrate on the computation of:
\begin{eqnarray}
\left|\beta(\omega_\in,\omega_\out)\right|^{2}
&=& \sum_{lm}\sum_{l^{\prime}m^{\prime}}
\left[
\beta_{lm,l^{\prime}m^{\prime}}(\omega_\in,\omega_\out)
\right]^2
\\
&=& 
R^2 
\left({
[
\{\gamma\left(\ngi;\ngo\right) - (\ngo)^2 \}
\omega_\out^2 
-
\{\gamma\left(\ngi;\ngo\right) - (\ngi)^2 \}
\omega_\in^2
]
\over 
\omega_\out+\omega_\in
}\right)^2 
\nonumber\\
&&
\qquad \times
\sum_{l=1}^\infty (2l+1) \;
 |\Xi_\in|^2  |\Xi_\out|^2 
 \left| A_{\nu}^\in \right|^{2}  
 \left| A_{\nu}^\out \right|^{2}
 \left[ 
        {W[J_{\nu}(\ngo\,\omega_\out r/c),
           J_{\nu}(\ngi\,\omega_\in r/c)]_{R}
        \over
        (\ngo\,\omega_\out)^2-(\ngi\,\omega_\in)^2} 
  \right]^2.
\label{E:b2}
\end{eqnarray}
(Note the symmetry under interchange of ``in'' and ``out''; moreover
$l=0$ is excluded since there is no monopole radiation for
electromagnetism.  Also, note that the refractive index of the
liquid in which the bubble is embedded shows up only indirectly: 
in the $A$ and $\Xi$ coefficients.) The above is a general result
applicable
to {\em any} dielectric sphere that undergoes sudden change in
refractive index. However, this expression is far too complex to
allow a practical analytical resolution of the general case.  For
the specific case of sonoluminescence, using our variant of the
dynamical Casimir effect, we shall show that the terms appearing
in it can be suitably approximated in such a way as to obtain a
tractable form that yields useful information about the main
predictions of this model.  We shall first consider the large volume
limit, which will allow us to compare this result to Schwinger's
calculation, and then develop some numerical approximations suitable
to estimating the predicted spectra for finite volume.

\section{The large $R$ limit}

The large $R$ limit is a reasonably good approximation to the
physical situation in sonoluminescence, since with our new
interpretation the radius of the light emitting region is about $500$
nm, which is somewhat larger than the short distance cutoff on the
wavelength ($\lambda_{\mathrm min} \approx 200 \; {\rm nm}$,
see~\cite{Companion}).  Independent of the issue of whether the large
$R$\, limit is a good fit to empirical reality, it is certainly useful
in its own right for giving us an analytically tractable qualitative
understanding of the physics of sudden dielectric changes in large
bubbles. (Compare this with the $k R \gg 1$ approximation invoked by
Eberlein.)

If $R$ is very large (but finite in order to avoid infra-red
divergences) then the ``in'' and the ``out'' modes can both be
described by ordinary Bessel functions
\begin{eqnarray}
G^\in(\ngi,\omega,r)&=& \Xi_\in \; 
J_{\nu}(\ngi\,\omega_\in r/c),\\
G^\out(\ngo,\omega,r)&=&\Xi_\out \; 
J_{\nu}(\ngo\,\omega_\out r/c).
\end{eqnarray}
We can now compute the Bogolubov
coefficients relating these
states
\begin{eqnarray}
\alpha_{ij}&=&(E^\out_{i},{E^\in_{j}})\\
&=& 
\Xi_\in^* \; \Xi_\out \;
{(\omega_\in+\omega_\out)\over c^2} 
e^{i(\omega_\out-\omega_\in)t}
\; \delta_{ll^{\prime}} 
\; \delta_{mm^{\prime}} 
\; \gamma\left(\ngi;\ngo\right)
\; \int J_{\nu}(\ngi\,\omega_\in r/c)
J_{\nu}(\ngo\,\omega_\out r/c) \; r\: dr\\
&=&
\Xi_\in^* \; \Xi_\out \;
(\omega_\in+\omega_\out)
e^{i(\omega_\out-\omega_\in)t} 
\; \delta_{ll^{\prime}} 
\; \delta_{mm^{\prime}}
\; \gamma\left(\ngi;\ngo\right)
\; \frac{\delta(\ngi\,\omega_\in-\ngo\,\omega_\out)}
{\ngi\,\omega_\in}\\
&=&
\Xi_\in^* \; \Xi_\out \; \gamma\left(\ngi;\ngo\right)
\left( {1\over\ngi} + {1\over\ngo}\right) 
e^{i(\omega_\out -\omega_\in)t} 
\; \delta_{ll^{\prime}}
\; \delta_{mm^{\prime}} 
\; \delta(\ngi\,\omega_\in-\ngo\,\omega_\out).
\end{eqnarray}
In terms of a plane-wave basis this is equivalent to
\begin{equation}
\alpha(\vec \kappa_\in,\vec \kappa_\out) = 
\gamma\left(\ngi;\ngo\right) \;
\left( {1\over\ngi} + {1\over\ngo}\right) \;
e^{i(\omega_\out -\omega_\in)t} \;
\delta^3(\vec \kappa_\in - \vec \kappa_\out)
\end{equation}
Note that as $\ngo\to\ngi$ the $\alpha$ coefficient has the correct limit:
\begin{equation}
\alpha_{ij} \to 2 \; \Xi_\in^* \; \Xi_\out \;
\delta_{ll^{\prime}}
\; \delta_{mm^{\prime}} 
\; \delta(\omega_\in-\omega_\out).
\end{equation}
Recall that the $\Xi$ factors have been carefully chosen to make sure
that the above is simply a three-dimensional delta function in momentum
space, translated into the spherical-polar basis. That is, in terms of
a plane-wave basis the large $R$ limit (for $\ni=\no$) is 
\begin{equation}
\alpha(\vec \kappa_\in, \vec \kappa_\out) \to \delta^3(\vec \kappa_\in - \vec \kappa_\out).
\end{equation}
The computation for $\beta$ is analogous
\begin{eqnarray}
\beta_{ij}&=&({E^\out_{i}}^*,{E^\in_{j}})
\\
&=&  
\Xi_\in \; \Xi_\out \;
{(\omega_\in-\omega_\out)\over c^2} 
e^{i(\omega_\out+\omega_\in)t}
\; \delta_{ll^{\prime}} 
\; \delta_{m,-m^{\prime}} 
\; \gamma\left(\ngi;\ngo\right)
\; \int J_{\nu}(\ngi\,\omega_\in r/c)
J_{\nu}(\ngo\,\omega_\out r/c) \; r\: dr\\
&=&
\Xi_\in \; \Xi_\out \;
(\omega_\in-\omega_\out)
e^{i(\omega_\out+\omega_\in)t} 
\; \delta_{ll^{\prime}} 
\; \delta_{m,-m^{\prime}}
\; \gamma\left(\ngi;\ngo\right)
\; \frac{\delta(\ngi\,\omega_\in-\ngo\,\omega_\out)}
{\ngi\,\omega_\in}\\
&=&
\Xi_\in \; \Xi_\out \;
\gamma\left(\ngi;\ngo\right)
\; \left( {1\over\ngi} - {1\over\ngo}\right) 
e^{i (\omega_\in+\omega_\out) t} 
\; \delta_{ll^{\prime}}
\; \delta_{m,-m^{\prime}} 
\; \delta(\ngi\,\omega_\in-\ngo\,\omega_\out).
\end{eqnarray}
In terms of a plane-wave basis this is equivalent to
\begin{equation}
\beta(\vec \kappa_i, \vec \kappa_j) =
\gamma\left(\ngi;\ngo\right) \; 
\left( {1\over\ngi} - {1\over\ngo}\right) 
e^{i (\omega_\in+\omega_\out) t} \;
\delta^3(\vec \kappa_\in + \vec \kappa_\out).
\end{equation}
Comparing this result with the independent calculation of the
companion paper~\cite{Companion} finally fixes the overall
normalization of $\gamma\left(\ngi;\ngo\right)$:
\begin{equation}
\gamma\left(\ngi;\ngo\right) = \ngi \; \ngo.
\end{equation}
Once this normalization is fixed, the large $R$ limit of our finite
volume calculation (which we have carried out only in the sudden
approximation) exactly reproduces the sudden limit of the homogeneous
dielectric calculation carried out in~\cite{Companion}. Readers are
referred to that paper for the corresponding numerical estimates of
the quantity and spectrum of the emitted photons.

We mention that with considerably more brute force analysis the same
results as above can also be obtained directly from equation
(\ref{E:b2}) by formally taking the $R \to \infty$ limit and using the
asymptotic formulae for the Bessel functions.

To estimate the possible significance of finite volume effects note
that in our new version of Schwinger's model we have $R\approx
R_{\mathrm light-emitting-region} \approx 500 \; \hbox{nm}$ and take
$K\approx 2\pi/(200\; {\mathrm nm})$ so that $KR\approx 5\pi \approx
15$.  (Which is why we expect that using the large R limit is a
tolerably good approximation, in addition to being much clearer for
gaining qualitative understanding). To get about one million photons
we need, for instance, $\ngi\approx 1$ and $\ngo\approx 12$, or
$\ngi\approx 2\times10^4$ and $\ngo \approx 1$, or even $\ngi \approx
71$ with $\ngo \approx 25$~\cite{Companion}.  Note that the estimated
values of $\ngo$ are extremely sensitive to the precise choice of
$\gamma$ and the high-frequency cutoff, and that the approximations
used in taking the large $R$ limit are at this stage uncontrolled.

\section{Behaviour for finite radius: Numerical analysis}

We now turn to the study of the predictions of the model in the
case of finite radius. Unfortunately this cannot be done 
analytically due to the wild behaviour of the pseudo--Wronskian
of the Bessel functions.  Nevertheless with some ingenuity,
and a detailed study of the different parts of the Bogolubov
coefficient, we are led to some reasonable approximations
that allow a clear description of the photon spectrum predicted
by the model.

\subsection{The A factor}

The $A_{\nu}$, $B_{\nu}$, and $C_{\nu}$ factors can be obtained
by a two step calculation. First one must solve the system
(\ref{E:coef}) by expressing $B$ and $C$ as functions of $A$. Then
one can fix $A$ by requiring $B^{2}+C^{2}=1$, a condition which
comes from the asymptotic behaviour of the Bessel functions.
Following this procedure, and again suppressing factors of $c$ for
notational convenience, we find that for the ``in'' coefficients
\begin{eqnarray}
A_{\nu}^\in 
&=& 
\left.
\frac{W[J_{\nu}(\nl\, \omega_\in r), 
        N_{\nu}(\nl\, \omega_\in r)]}
{\sqrt{W[ J_{\nu}(\ngi\, \omega_\in r), 
          N_{\nu}(\nl\, \omega_\in r)]^2+ 
       W[ J_{\nu}(\ngi\, \omega_\in r), 
          J_{\nu}(\nl\, \omega_\in r)]^2}}
\right|_{R},
\\ 
B_{\nu}^\in 
&=& 
\left. A_{\nu}^\in  
\frac{W[J_{\nu}(\ngi\,\omega_\in r), 
        N_{\nu}(\nl\, \omega_\in r)]}
{W[ J_{\nu}(\nl\, \omega_\in r), 
    N_{\nu}(\nl\, \omega_\in r) ]}
\right|_{R},
\\
C_{\nu}^\in 
&=& 
\left.
A_{\nu}^\in  
\frac{W[J_{\nu}(\nl\, \omega_\in r), 
      J_{\nu}(\ngi\, \omega_\in r)]}
{W[J_{\nu}(\nl\, \omega_\in r), 
   N_{\nu}(\nl\, \omega_\in r)]}
\right|_{R}.
\label{E:coef2}
\end{eqnarray}
We are mostly interested in the coefficient $A_{\nu}$. This can be
simplified by using a well known formula (cf. \cite{Abramowitz}, page
360 formula 9.1.16) for the (true) Wronskian of Bessel functions of the
first and second kind.
\begin{equation}
W_{true}[J_{\nu}(z), N_{\nu}(z)]=\frac{2}{\pi z}.
\end{equation} 
In our case, taking into account that for our pseudo--Wronskian the
derivatives are with respect to $r$ (not with respect to $z$), one
gets for the numerator of $A_{\nu}$:
\begin{equation}
W[J_{\nu}(\nl\, \omega_\in r), 
  N_{\nu}(\nl\, \omega_\in r)]_{R}
= \nl\, \omega_\in 
\frac{2}{\pi (\nl\, \omega_\in R)}=\frac{2}{\pi R}.
\end{equation}
Hence the $A_{\nu}$ factor can be written as
\begin{equation}
|A_{\nu}|^{2}= 
\frac{4/(\pi^2 R^2)}{\left. 
W[ J_{\nu}(\ngi\,\omega_\in r), 
   N_{\nu}(\nl\, \omega_\in r)]^2+
W[ J_{\nu}(\ngi\,\omega_\in r), 
   J_{\nu}(\nl\, \omega_\in r)]^2 
\right|_{R}}.
\end{equation}
Now adopt the notation $y = \ngi\, \omega_\in R/c$ and
$y_\liquid = \nl\, \omega_\in R/c = (\nl/\ngi) y$. Then
\begin{equation}
|A_{\nu}^\in(y,y_\liquid)|^{2} = 
{4/\pi^2 \over
  \left|
  \begin{array}{rr}
  J_{\nu}(y)&N_{\nu}(y_\liquid)\\
  y\;J^{\prime}_{\nu}(y) & y_\liquid\; 
                           N^{\prime}_{\nu}(y_\liquid)
  \end{array}
  \right|^{2}
+
  \left|
  \begin{array}{rr}
  J_{\nu}(y)&J_{\nu}(y_\liquid)\\
  y\;J^{\prime}_{\nu}(y) & y_\liquid\;
                           J^{\prime}_{\nu}(y_\liquid)
  \end{array}
  \right|^{2}
}, 
\end{equation}
where in this equation the primes now signify derivatives with
respect to the full arguments ($y$ or $y_\liquid$). A
similar formula holds of course for $A_\nu^\out$ in terms
of $x$ and $x_\liquid$. Using the standard identities
$ x J'_\nu(x) = x J_{\nu-1}(x) - \nu J_\nu(x)$ and $ x N'_\nu(x)
= x N_{\nu-1}(x) - \nu N_\nu(x)$, applying properties of the
determinant, and adopting the notation $\Ni = \nl/\ngi$, this can be
simplified to 
\begin{equation}
|A_{\nu}^\in(y,\Ni)|^{2} = 
{4/\pi^2 \over
  \left|
  \begin{array}{rr}
  J_{\nu}(y)&N_{\nu}(\Ni y)\\
  y\;J_{\nu-1}(y) & \Ni y\; N_{\nu-1}(\Ni y)
  \end{array}
  \right|^{2}
+
  \left|
  \begin{array}{rr}
  J_{\nu}(y)&J_{\nu}(\Ni y)\\
  y\;J_{\nu-1}(y) & \Ni y\; J_{\nu-1}(\Ni y)
  \end{array}
  \right|^{2}
}.
\end{equation}
By considering the small argument expansions for the Bessel
functions it is relatively easy to see that for small $y$ (holding
$\Ni$ fixed)
\begin{equation}
|A_{\nu}^\in(y\to 0,\Ni)|^{2} \to  (\Ni)^{2\nu} + O(y).
\end{equation}
On the other hand, for large values of the argument $y$ the asymptotic
forms of the Bessel functions can be used to demonstrate that 
\begin{equation}
|A_{\nu}^\in(y\to \infty,\Ni)|^{2}
\sim
{2\ngi\,\nl\over (\ngi)^2+\nl^2 + 
[\nl^2-(\ngi)^2]\sin(2 y  - \nu\pi)}.
\end{equation} 
Numerical plots of  $|A_\nu|^2$  show that it is an oscillating 
function of $y$ which rapidly reaches this asymptotic form.
The mean value for large arguments is simply:
\begin{equation}
|A_{\nu}^\in(y\to \infty,\Ni)|^2 \approx {1\over2\pi} 
\int_0^{2\pi} dz 
{2\ngi\,\nl\over (\ngi)^2+\nl^2 + [\nl^2-(\ngi)^2]\sin(z)} = 1.
\label{E:app}
\end{equation}
While these results are general, for the particular application to SL
that we have in mind it is the small $y$ behaviour that is most
relevant. Also, keep in mind that this large $y$ asymptotic formula
holds for $y$ large but ignoring dispersive effects (that is, assuming
a frequency independent index of refraction). If we model dispersive
effects by a Schwinger-like cutoff where the refractive index drops to
unity (see below) then above the cutoff we will have $A_\nu \equiv 1$
holding as an identity.

\subsection{The Pseudo--Wronskian}

Use the simplified notation in which $x = \ngo\, \omega_\out R/c$, $y
= \ngi\, \omega_\in R/c$. In these dimensionless quantities, after
making explicit the dependence on $R$ and $c$, and inserting the
particular choice of $\gamma$ motivated by the large-$R$ limit,
equation (\ref{E:b2}) takes the form:
\begin{equation}
\label{E:b2-2}
\left|\beta(x,y)\right|^{2}=\frac{R^2}{c^2} 
\left(\ngo-\ngi\right)^2 \;
 |\Xi_\in|^2 \; |\Xi_\out|^2 \;
\left(
{\ngi \; x^2 + \ngo \; y^2
\over
\ngi \; x + \ngo \; y }
\right)^2
F(x,y).
\end{equation}
Here $F(x,y)$ is shorthand for the function
\begin{equation}
F(x,y) = \sum_{l=1}^\infty (2l+1) \; 
|A_l{}^\in|^2 \; |A_l{}^\out|^2
{ 
  \left|
  \begin{array}{rr}
  J_{\nu}(x)&J_{\nu}(y)\\
  x\;J^{\prime}_{\nu}(x) & y\;J^{\prime}_{\nu}(y)
  \end{array}
  \right|^{2}
\over
(x^2-y^2)^2
},
\label{E:bint}
\end{equation}
where in this equation the primes now signify derivatives with respect
to the full arguments ($x$ or $y$). It is convenient to define a
dimensionless Bogolubov coefficient, and a dimensionless spectrum, by
taking
\begin{equation}
|\beta(x,y)|^2 = {R^2\over c^2} \; |\beta_0(x,y)|^2,
\end{equation}
so that
\begin{equation}
\label{E:N-spectrum-dimensionless}
{dN(x)\over dx} 
= {1\over 4\;\ngi\ngo} \int_0^\infty dy \;
{|\beta_0(x,y)|^2 \over |\Xi_\in|^2 \; |\Xi_\out|^2 }.
\end{equation}
The total number of photons is then
\begin{equation}
\label{E:N-dimensionless}
N 
= {1\over 4\;\ngi\ngo} \int _0^\infty dx \int_0^\infty dy \;
{|\beta_0(x,y)|^2 \over |\Xi_\in|^2 \; |\Xi_\out|^2 }.
\end{equation}
The total energy emitted is given by a very similar formula\footnote{%
For a flash occurring at minimum radius $\hbar c/R \approx 0.4 \;
\rm{eV}.$}
%
\begin{equation}
\label{E:E-dimensionless}
E
= {\hbar c \over R \ngo} \; {1\over 4\;\ngi\ngo} 
\int _0^\infty dx \int_0^\infty dy \; x\;
{|\beta_0(x,y)|^2 \over |\Xi_\in|^2 \; |\Xi_\out|^2 }.
\end{equation}
In order to proceed in our analysis we need now to perform the
summation over angular momentum.  Although the infinite sum is
analytically intractable, we can easily demonstrate that it is
convergent and can physically argue that the lowest angular momentum
modes will dominate the sum.  Consider the large order expansion ($\nu
\gg x$ at fixed $x$) of the Bessel functions. In this limit one gets
\cite{Jeffrey}:
\begin{equation}
\label{E:asymp}
J_{\nu}(x) 
\sim {\frac{1}{\sqrt{2\pi \nu}}}\left(\frac{e x}{2\nu}\right)^{\nu}
\end{equation}
This can be used to obtain the asymptotic form of the 
pseudo--Wronskian appearing in equation  (\ref{E:bint}).
\begin{eqnarray}
\tilde W_\nu(x,y)&\equiv&\left|
\begin{array}{rr}
J_{\nu}(x)&J_{\nu}(y)\\
x\;J^{\prime}_{\nu}(x) & y\;J^{\prime}_{\nu}(y)
\end{array}
\right|\\
&=&-\left|
\begin{array}{rr}
J_{\nu}(x)&J_{\nu}(y)\\
x\;J_{\nu+1}(x) & y\;J_{\nu+1}(y)
\end{array}
\right|\\
&\sim& \frac{(x^2-y^2)}{2\pi (\nu)^{1/2} (\nu+1)^{3/2}} 
\left(\frac{xy}{\nu(\nu+1)}\right)^{\nu}  
\left(\frac{e}{2}\right)^{2\nu+1}.
\end{eqnarray}
where we have used the standard recursion relation for the Bessel
functions $J^{\prime}_{\nu}(z)=\nu J_{\nu}(z)-z J_{\nu+1}(z)$.
This indicates that the sum over $\nu$ is convergent: the terms
for which $(xy/\nu^{2})\leq 1$ are suppressed. Whatever the values
of $x$ and $y$ are, for sufficiently large angular momenta this
asymptotic form guarantees the convergence of the sum over angular
momenta.

Everything so far has been predicated on the absence of dispersion:
the refractive index is independent of frequency. In real physical
materials the refractive index is known to fall to unity at high
enough energies. (Sufficiently high energy photons ``see'' a vacuum
inhabited by effectively-free isolated charged particles. The manner
in which the refractive index approaches unity is governed by the
plasma frequency, and the location of this physical cutoff is governed
by the resonances present in the atomic structure of the atoms.) This
situation is far too complex to be modelled in detail, but it is easy
to see that an upper bound on emitted photon energies implies an upper
bound on the allowed angular momentum modes: Basically, if one
supposes the photons to be produced inside or at most on the surface
of the light emitting region, then the upper limit for the angular
momentum (as seen at spatial infinity) will be attained by photons
emitted tangentially from the edge of the light emitting region: this
maximal angular momentum is the product of the radius of the light
emitting region times the maximum observed ``out'' momentum.  Then one
gets:
\begin{equation}
l_{\mathrm max}^\outside
= { (\hbar K_{observed}) \times R \over \hbar} 
= R K_{observed}.
\label{E:lmax}
\end{equation}
For sonoluminescence $K_{observed}$ is of order $2\pi/(200\; {\rm
nm})$. Since the light emitting region is known to be approximately
$500 \; {\rm nm}$ wide we shall be most interested in the case $K
R\approx 5\pi \approx 15$, with a corresponding maximum angular
momentum $l_{\mathrm max}$ approximately $15$.  Under these
conditions, the bulk of the radiation will be into the lowest allowed
angular momentum modes. The precise value of the angular momentum
cutoff $l_{\mathrm max}$ is sensitive to the details of both the
frequency cutoff in refractive index, and the size of the light
emitting region. For instance, in some of Schwinger's papers he took
$K \approx 2\pi/(400\; {\rm nm})$ in which case (taking again $R
\approx 400\; {\rm nm}$) $l_{\mathrm max} \approx 5\pi/2 \approx 7$.
Whatever ones views as to the precise value of this cutoff it is clear
that the emitted radiation is limited to low angular momenta.

A subtlety is that this is the angular momentum as measured at spatial
infinity (in the ambient liquid---water). This is not the same as the
angular momentum the photons have while they are inside the bubble
(since it is frequency, not wavenumber, that is conserved when photons
cross a timelike interface [spacelike normal]).\footnote{%
Contrast this to a spacelike interface (timelike normal; sudden
temporal change in the refractive index) for which it is the
wavenumber, not the frequency, that is conserved across the
interface. During photon {\em production} we are dealing with a
spacelike interface, whereas when the photons {\em escape} from the
gas bubble we are dealing with a timelike interface. }
Taking this into account
\begin{equation}
l_{\mathrm max}^\inside = {\ngo\over\nl} \; l_{\mathrm max}^\outside
= {\ngo\over\nl}\;  R K_{observed} \approx  {\ngo\over\nl}\; 15.
\label{E:lmax-2}
\end{equation}
We now discuss how to take this observationally based cutoff in
angular momentum and translate it into a statement about the cutoff
in the refractive index.

\bigskip

\subsection{Implementation of the cutoff}

If we adopt a Schwinger-like momentum-space cutoff in the refractive
index, then because we have defined the variables $x$ and $y$ partly
in terms of the refractive index, we must carefully assess the meaning
of these variables. In terms of momenta, Schwinger's cutoff is
\begin{equation}
\ni(\kappa) = \ni \; \Theta(K_\in -\kappa) + 1 \; \Theta(\kappa-K_\in),
\end{equation}
\begin{equation}
\no(\kappa) = \no \; \Theta(K_\out -\kappa) + 1 \; \Theta(\kappa-K_\out),
\end{equation}
This implies that the photon dispersion relation $\omega(\kappa)$ has a
kink at $\kappa=K$, and that we can write
\begin{equation}
\omega_\in(\kappa) = {c \kappa\over \ni}\; \Theta(K_\in -\kappa) + 
\left({c K_\in\over \ni} + c (\kappa-K_\in) \right) \; \Theta(\kappa-K_\in),
\end{equation}
\begin{equation}
\omega_\out(\kappa) = {c \kappa\over \no}\; \Theta(K_\out -\kappa) + 
\left({c K_\out\over \no} + c (\kappa-K_\out) \right) \; \Theta(\kappa-K_\out).
\end{equation}
Finally, the variables $x$ and $y$ generalize (actually, simplify) to
\begin{equation}
x = \kappa_\out R/c; \qquad y = \kappa_\in R/c,
\end{equation}
so that
\begin{equation}
\label{E:ny}
\ni(y) = \ni \; \Theta(y_* -y) + 1 \; \Theta(y-y_*),
\end{equation}
\begin{equation}
\label{E:nx}
\no(x) = \no \; \Theta(x_*-x) + 1 \; \Theta(x-x_*),
\end{equation}
{\tt where $x_* \equiv K_\out R/c;\ y_* \equiv K_\in R/c$.} 
Now all these changes do not affect $F(x,y)$, which is why we defined
it the way we did, but they do affect the prefactors appearing in
equation (\ref{E:b2-2}).  An immediate consequence is that the $(x,y)$
plane naturally separates into four regions and that $|\beta(x,y)|^2 =
0$ in the region $x>x_*$ and $y>y_*$. We shall soon see that the two
``tail'' regions $(x<x_*,y>y_*)$ and $(x>x_*,y<y_*)$ are relatively
uninteresting, and that the bulk of the contribution to the emission
spectrum comes from the region $(x<x_*,y<y_*)$.\footnote{%
If one is too enthusiastic about adopting the sudden approximation then
the integral over these tail regions will be divergent. This, however,
is not a physical divergence, but is instead a purely mathematical
artifact of taking the sudden approximation all the way out to
infinite frequency. The integral over these two tail regions is in
fact cut off by the fact that for high enough frequency the sudden
approximation breaks down. As a practical matter we have found that
the numerical contribution from these tail regions are small.}

In the infinite volume limit this is an exact statement, since in that
limit one can show (see~\cite{Companion} and the discussion below)
that $F(x,y)\to G(x)\;\delta(x-y)$ so that the support of the spectral
integral is exactly the line segment $x=y$ with $x=y \leq
\min\{x_*,y_*\}$.

Finally, when it comes to choosing specific values for $x_*$ and
$y_*$, we use the fact that the variables $x$ and $y$ are related to
the angular momentum cutoff discussed in the previous subsection to
set
\begin{equation}
x_* = y_* =  {\ngo\over\nl}\; 15.
\end{equation}
%

\subsection{Working along the diagonal}

To study in more detail the behaviour of the function $F(x,y)$ when
higher angular momentum modes are retained one can perform a Taylor
expansion of $F(x,y)$ around $x=y$.
\begin{eqnarray}
\lim_{x\rightarrow y} \frac{\tilde W_{\nu}(x,y)}{(x-y)} 
&\equiv& \lim_{x\rightarrow y}
\frac{ 
\left|\begin{array}{rr}
J_{\nu}(x)&J_{\nu}(y)\\
x\;J^{\prime}_{\nu}(x) & y\;J^{\prime}_{\nu}(y)
\end{array}
\right|}
{(x-y)}\\
&=& \lim_{x\rightarrow y}
\frac{
\left|\begin{array}{ll}
J_{\nu}(x)& \hphantom{x\;} J_{\nu}(x)+(x-y)J^{\prime}_{\nu}(x)\\
x\;J^{\prime}_{\nu}(x) & 
x\;J^{\prime}_{\nu}(x)+(x-y)
[J^{\prime}_{\nu}(x)+x\;J^{\prime\prime}_{\nu}(x)] 
\end{array} 
\right|}   
{(x-y)}\\
&=& 
\left|\begin{array}{rl}
J_{\nu}(x)&J^{\prime}_{\nu}(x)\\
x\;J^{\prime}_{\nu}(x) & 
J^{\prime}_{\nu}(x)+x\;J^{\prime\prime}_{\nu}(x)
\end{array} 
\right| \\
&=&
J_{\nu}(x)[J^{\prime}_{\nu}(x)+x\;J^{\prime\prime}_{\nu}(x)]-
x\;{J^{\prime}_{\nu}(x)}^{2}.
\end{eqnarray}
The derivatives can be eliminated by using the well known recursion 
relations.
\begin{eqnarray} 
\lim_{x\rightarrow y} \frac{\tilde W_{\nu}(x,y)}{(x-y)}
&=&
J_{\nu}(x)
\left[\frac{(\nu^{2}-x^{2})}{x}\right]
-x\;\left[\frac{\nu}{x} J_{\nu}(x)-J_{\nu+1}(x)\right]^{2}\\
&=& 
2 \nu 
J_{\nu}(x)J_{\nu+1}(x)-x\;\left[J_{\nu}^{2}(x)+J_{\nu+1}^{2}(x)\right].
\label{E:diatrm} 
\end{eqnarray} 
For sake of simplicity we shall use an equivalent form of equation
(\ref{E:diatrm}) where lower order Bessel functions appear
\begin{equation}
\lim_{x\rightarrow y} \frac{\tilde W_{\nu}(x,y)}{(x-y)}=2 \nu 
J_{\nu}(x)J_{\nu-1}(x)-x\;\left[J_{\nu}^{2}(x)+J_{\nu-1}^{2}(x)\right].
\label{E:dia}
\end{equation} 
This result shows that, as expected, each term of $F(x,x)$ is 
finite along the diagonal and equal to zero at $x=y=0$.
Moreover 
\begin{equation}
D(x)\equiv F(x,x)=\sum_{l=1}^{\infty}(2l+1) 
{\left\{
(2l+1) J_{l+1/2}(x) J_{l-1/2}(x) 
-x \; \left[J_{l+1/2}^{2}(x)+J_{l-1/2}^{2}(x)\right]
\right\}^2
\over
4x^{2}}. 
\end{equation} 
This sum can easily be checked to be convergent for fixed $x$. [Use
equation (\ref{E:asymp}).] With a little more work it can be shown that
\[
\lim_{x\to\infty} D(x) = {1\over2\pi^2}.
\]
The truncated function obtained after summation over the first few
terms (say the first ten or so terms) is a long and messy combination
of trigonometric functions that can however be easily plotted and
approximated in the range of interest.  A semi-analytical study led us
to the approximate form of $D(x)$
\begin{equation}
D(x) \approx \frac{1}{2\pi^{2}}
\frac{x^{6}}{250+x^{6}}.
\end{equation}
A confrontation between the two curves in the range of interest is given 
in the figure below.

\begin{figure}[htb]
\vbox{ 
\hfil
\scalebox{0.66}{\rotatebox{270}{\includegraphics{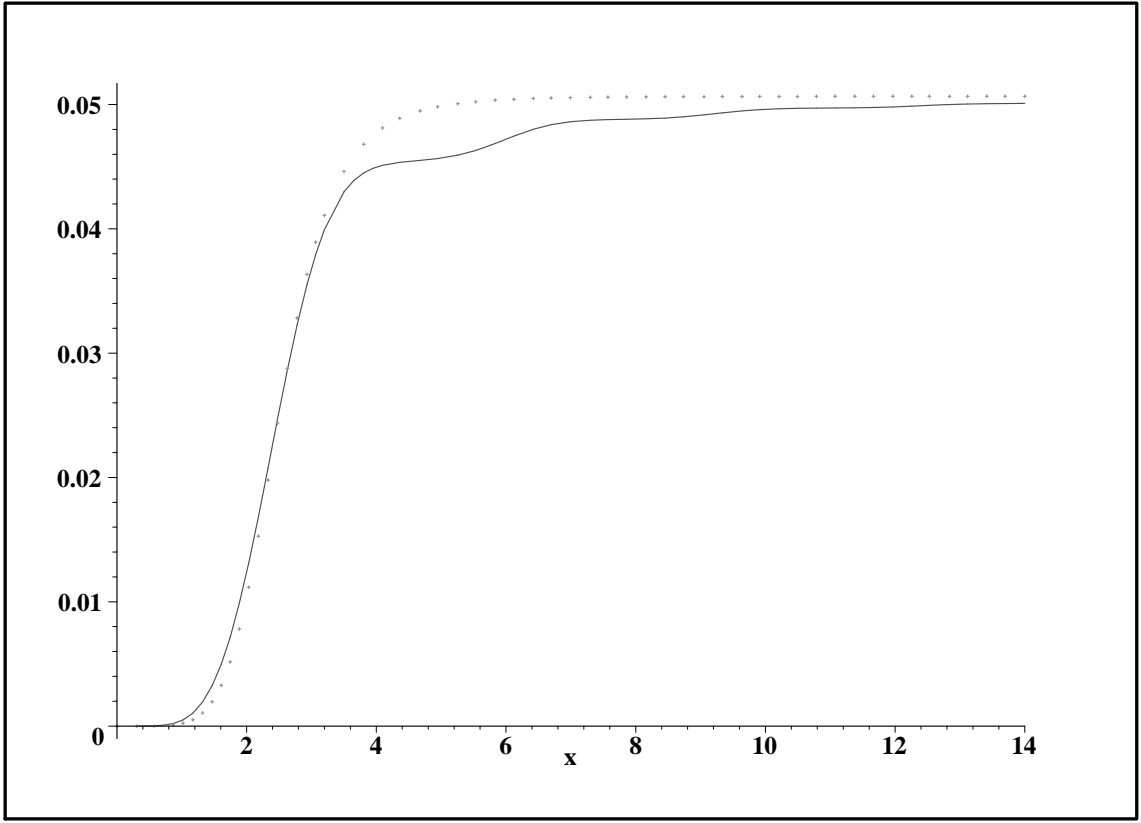}}}
\hfil 
}
\bigskip
\caption{
Plot of the exact $D(x)$ against its approximated form (dotted curve)
in the range $0<x<14$
}  
\label{F:diagonal}
\end{figure}

\subsection{The factorization approximation}

To numerically perform the integrals needed to do obtain the spectrum
it is useful to note the approximate factorization property
\begin{equation}
F(x,y) \approx F\left({x+y\over2},{x+y\over2}\right) \; 
G\left(\frac{x-y}{2}\right). 
\end{equation}
That is: to a good approximation $F(x,y)$ is given by its value
along the nearest part of the diagonal, multiplied by a universal
function of the distance away from the diagonal. A little experimental
curve fitting is actually enough to show that to a good approximation
\begin{equation}
F(x,y) \approx  D\left({x+y\over2}\right) \; 
{\sin^2(3[x-y]/4)\over (3[x-y]/4)^2} .
\end{equation}
{From} the plot we show below it is easy to check that the function
$F(x,y)$ is quite well approximated by this factorized form. We feel
important to stress that this is approximation is based on numerical
experimentation, and is not an analytically-driven approximation.  (In
the infinite volume case we know that $F(x,y) \to (constant) \times
\delta(x-y)$. The effect of finite volume is effectively to ``smear
out'' the delta function. In this regard, it is interesting to
observe that the combination $\sin^2(x)/(\pi x^2)$ is one of the
standard approximations to the delta function.)  Our approximation is
quite good everywhere except for values of $x$ and $y$ near the origin
(less than 1) where the contribution of the function to the integral
is very small.

\begin{figure}
\vbox{
\hfil
\scalebox{0.66}{\rotatebox{270}{\includegraphics{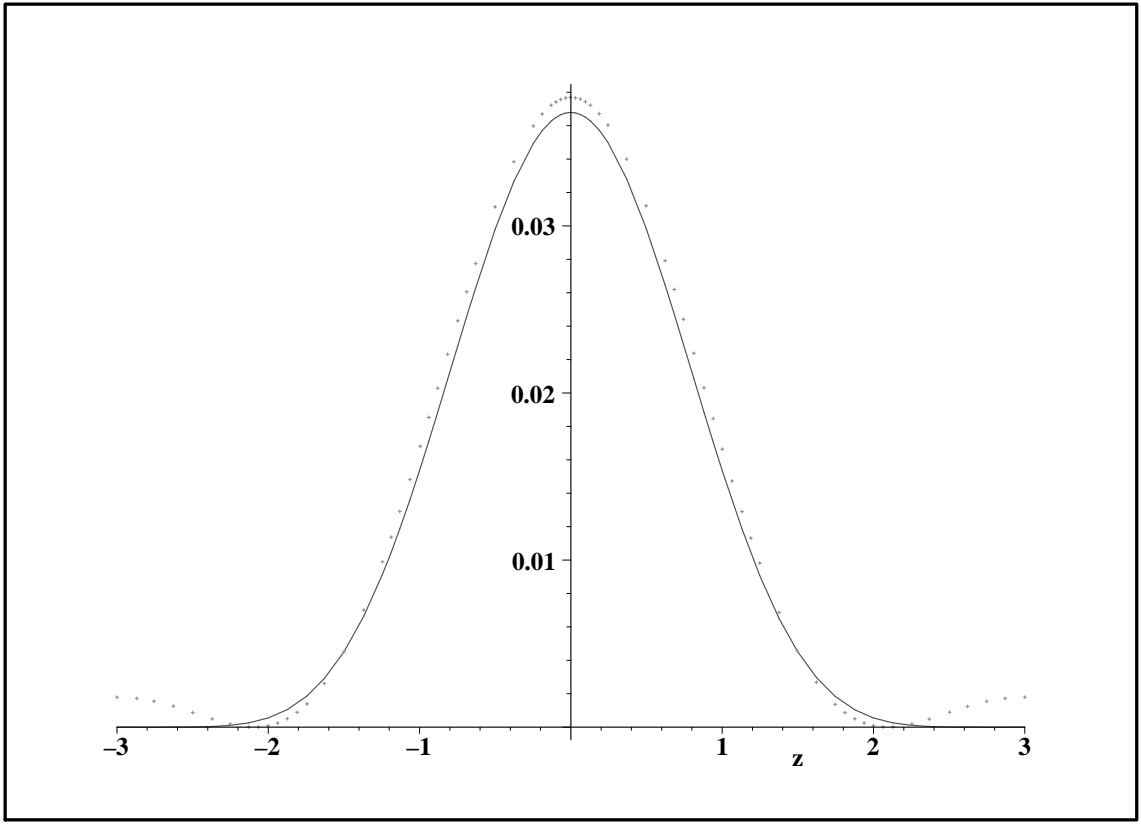}}}
\hfil
}
\bigskip
\caption{%
Transverse fit: An orthogonal slice of $F(x,y)$ intersecting the
diagonal at $(x,y)=(3,3)$. Here $F(3+z,3-z)$ is plotted in comparison
with $[\sin^2(3 z/2)]/(3 z/2)^2$. The solid line is the function, and
the dotted line the analytic approximation
}
\label{F:transverse}
\end{figure}

\subsection{The spectrum: numerical evaluation}

We have now transformed the function $F(x,y)$ into an easy to handle
product of two functions
\begin{equation}
F(x,y) \approx  
\frac{1}{2\pi^{2}}
\frac{(x+y)^{6}}{16000+(x+y)^{6}}
{\sin^2(3[x-y]/4)\over(3[x-y]/4)^2}. 
\label{fapp}
\end{equation}
We exhibit tridimensional graphs for both the exact (apart from the
approximation of truncating the sum at a finite $l$) and approximate
forms of the function $F(x,y)$.  We have chosen the case of $R=500 \;
{\rm nm}$ (corresponding to $y_{*}=15 \;\ngo/\nl$ as previously
explained).
%
\begin{figure}[htb]
\vbox{
\hfil
\scalebox{0.50}{\rotatebox{270}{\includegraphics{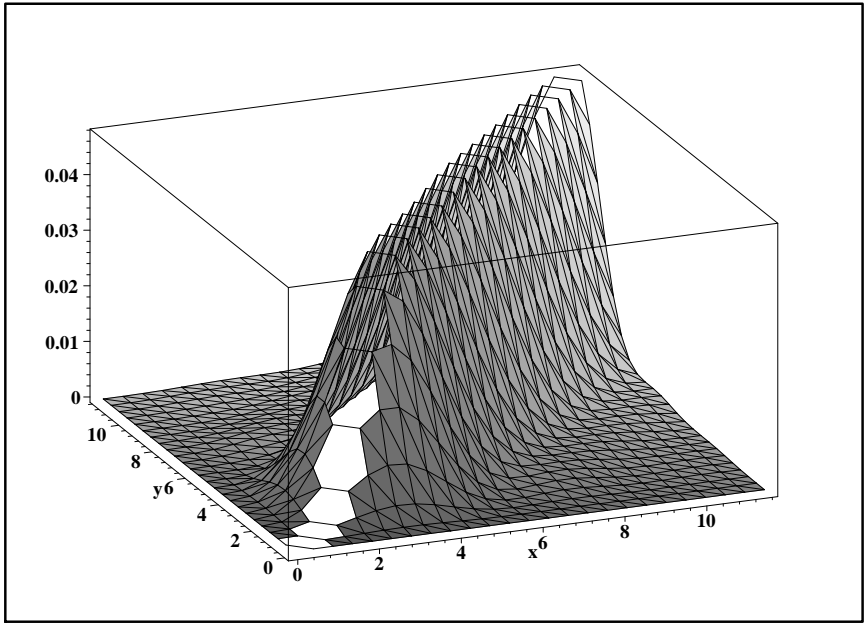}}}
\hfil
}
\bigskip
\caption{%
Plot of the exact $F(x,y)$ in the range $0<x<12$, $0<y<12$. The jagged
behaviour along the diagonal is a numerical artifact, as the function
is known to be smooth there.
}
\label{F:exact}
\end{figure}

\begin{figure}[htb]
\vbox{
\hfil
\scalebox{0.50}{\rotatebox{270}{\includegraphics{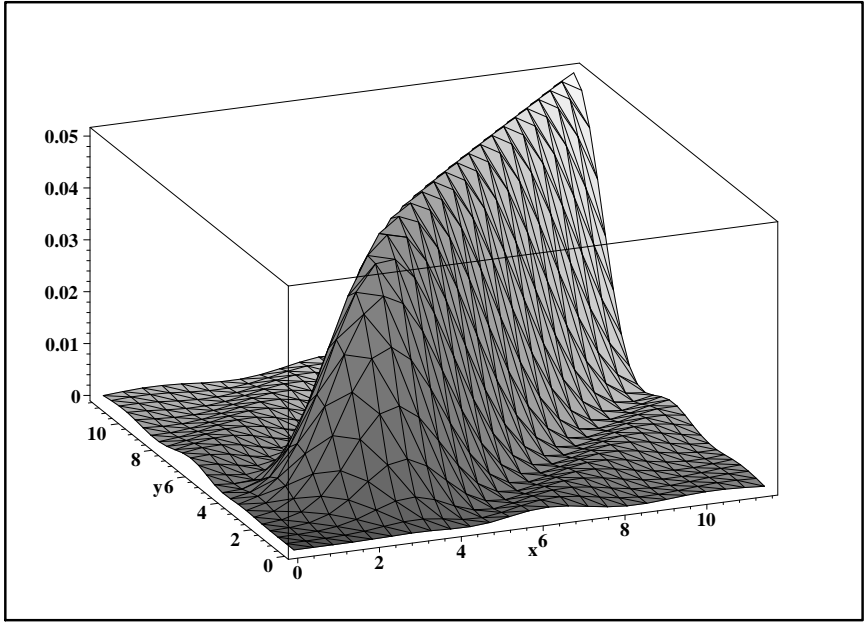}}}
\hfil
}
\bigskip
\caption{%
Plot of the approximated $F(x,y)$ in the range $0<x<12$,
$0<y<12$
}
\label{F:factorized}
\end{figure}

%
The dimensionless spectrum, based on equations (\ref{E:b2}) and
(\ref{E:b2-2}), is
\begin{equation}
\label{E:dimensionless-spectrum-2}
{dN\over dx} 
=\frac{(\ngi-\ngo)^2}{2 \;\ngi \ngo} 
\int_0^\infty
\left( \frac{\ngi \, x^2+\ngo \, y^2}{\ngi \, x+\ngo \, y} \right)^2 
D\left({x+y\over2}\right) 
{\sin^2(3[x-y]/4)\over(3[x-y]/4)^2} dy,
\end{equation}
where $\ngo(x)$ and $\ngi(y)$ are now the appropriate functions of $x$
and $y$ (See equations (\ref{E:nx}) and (\ref{E:ny})). We have also
manually inserted a factor $2$ to account for the photon polarizations.

As a consistency check, the infinite volume limit is equivalent to
making the formal replacements
\begin{equation}
{\sin^2(3[x-y]/4)\over(3[x-y]/4)^2} \to {4\pi\over3} \delta(x-y),
\end{equation}
and
\begin{equation}
D\left({x+y\over2}\right) \to {1\over2\pi^2}.
\end{equation}
The first replacement can be formally justified as follows. It is known 
that a sequence of smooth functions approximating the delta function is 
given by
\begin{equation}
f_{s}(x)=\frac{1}{s\; \pi}\; \frac{\sin^2 (s\; x)}{x^2};
\end{equation}
indeed, one get 
\begin{equation}
\lim_{s\to \infty}\; f_{s}(x)=\delta(x).
\end{equation}
Then, it is straightforward to show that
\begin{eqnarray}
{\sin^2(3[x-y]/4)\over(3[x-y]/4)^2} &=& 
{\sin^2(R \; 3[\ngi \oi-\ngo \oo]/(4 c))
\over(R \; 3[\ngi \oi-\ngo \oo]/(4 c))^2}
\nonumber\\
&\to& \frac{\pi}{R}\; \delta ( \pi[\ngi \oi-\ngo \oo]/(4 c))
\nonumber\\
&=& {4\pi\over3} \; \delta(x-y).
\end{eqnarray}
Doing so, equation (\ref{E:dimensionless-spectrum-2}) reduces to the
spectrum obtained for homogeneous dielectrics in~\cite{Companion}.
Indeed 
\begin{equation}
{dN\over dx} = 
{1\over3\pi} \frac{(\ngi-\ngo)^2}{\ngi \ngo}  \; x^2 \; \Theta(x_* - x).
\end{equation}
With these consistency checks out of the way, it is now possible to
perform the integral with respect to $y$ to estimate the spectrum for
finite volume, and similarly to perform appropriate double integrals
with respect to $x$ and $y$ to estimate both total photon production
and average photon energy.  In our companion paper~\cite{Companion} we
showed that in the infinite volume limit there were two continuous
branches of values for $\ngi$ and $\ngo$ that led to approximately one
million emitted photons with an average photon energy of $3/4$ the
cutoff energy. If we now place the same values of refractive index
into the formula (\ref{E:dimensionless-spectrum-2}) derived above,
numerical integration again yields approximately one million photons
with an average photon energy of $3/4$ times the cutoff energy. The
total number of photons is changed by at worst a few percent, while the
average photon energy is almost unaffected. (Some specific sample
values are reported in Table I.) The basic result is this: as
expected~\cite{Companion}, finite volume effects do not greatly modify
the results estimated by using the infinite volume limit. Note that
$\hbar \Omega_{\mathrm max}$ is approximately $4$ eV, so that average
photon energy in this crude model is about $3$ eV.

\begin{center}
\bigskip
\begin{tabular}{|c|c|c|c|}
\hline
$\;\ngi\;$ & $\;\ngo\;$  & Number of photons & 
$\langle E \rangle/\hbar \Omega_{\mathrm max}$ \\
\hline
\hline
$2\times10^4$ & $1$  & $1.06\times 10^6$ & $0.803$ \\
\hline
$71  $   & $25$ & $1.00\times 10^6$ & $0.750$ \\
\hline
$68  $   & $34$ & $1.06\times 10^6$ & $0.751$ \\
\hline
$9$   & $25$ & $0.955\times 10^6$ & $0.750$ \\
\hline
$1$      & $12$ & $0.98\times 10^6$ & $0.765$ \\
\hline
\end{tabular}
\medskip
\center{Table I: Some typical cases.}
\bigskip
\label{T:table}
\end{center}

In addition, for the specific case $\ngi=2\times10^4$, $\ngo=1$, we
have calculated and plotted the form of the spectrum. We find that the
major result of including finite volume effects is to smear out the
otherwise sharp cutoff coming from Schwinger's step-function model for
the refractive index. Other choices of refractive index lead to
qualitatively similar spectra.  These results are in reasonable
agreement (given the simplicity of the present model) with
experimental data.
%
\begin{figure}[htb]
\vbox{
\hfil
\scalebox{0.66}{\rotatebox{270}{\includegraphics{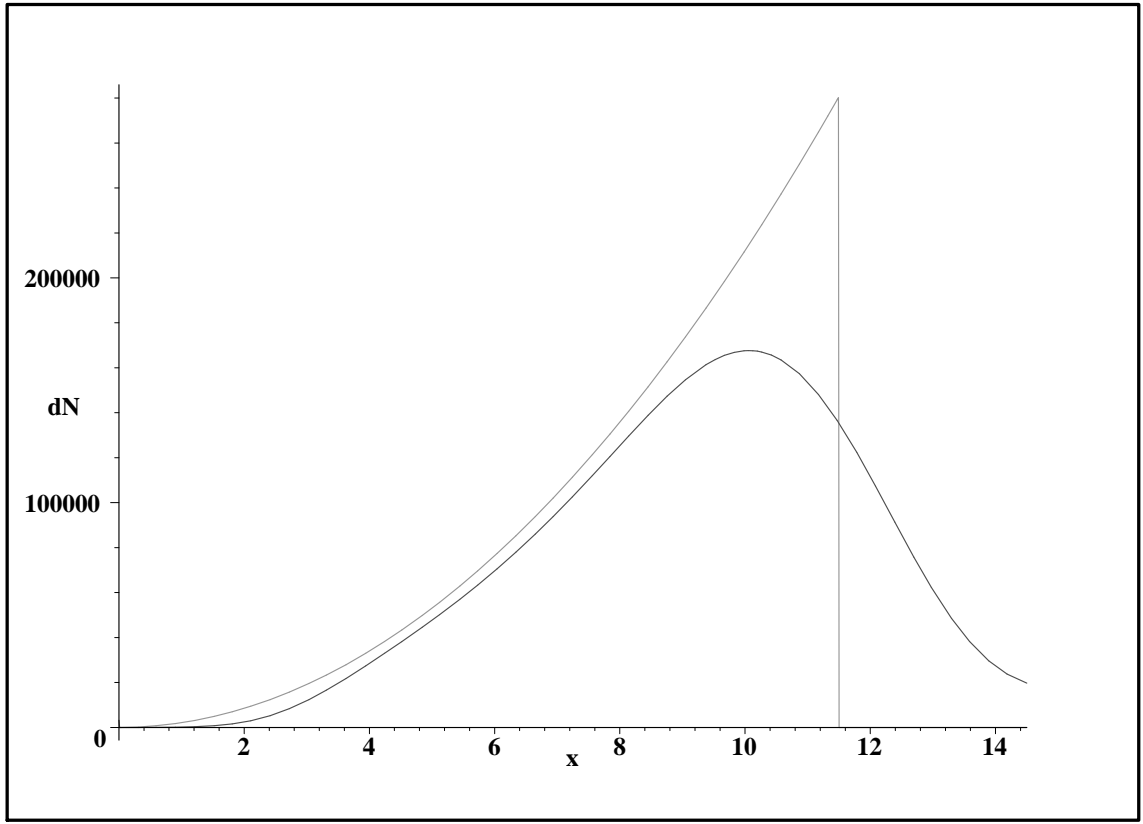}}}
\hfil
}
\bigskip
\caption{%
Spectrum $dN/dx$ obtained by integrating the approximated Bogolubov
coefficient.  We integrate from $y=0$ to $y_*=11.5$ and plot the
resulting spectrum from $x=0$ to $x=14.5$.
For $\no=1$ and $R=500$nm the relation between the non-dimensional
quantitiy $x$ and the frequency $\nu$ is $x\sim \nu \cdot 10.5 \cdot
10^{-15}$s. So $x\approx 11.5$ corresponds to $\nu \approx  1.1$ PHz.
The curve with the sharp
cutoff is the infinite volume approximation. Finite volume effects
tend to smear out the sharp discontinuity, but do not greatly affect
the total number of photons emitted.
} 
\label{F:spectrum} 
\end{figure}
%

\section{Discussion}

The present paper presents calculations of the Bogolubov coefficients
relating the two QED vacuum states appropriate to changes in the
refractive index of a dielectric bubble.  We have verified by explicit
computation that photons are produced by rapid changes in the
refractive index, and are in agreement with Schwinger in that QED
vacuum effects remain a viable candidate for explaining SL. However,
some details of the particular model considered in the present paper
are somewhat different from that originally envisaged by
Schwinger. Based largely on the fact that efficient photon production
requires timescales of the order of femtoseconds we were led to
consider rapid changes in the refractive index as the gas bubble
bounces off the van der Waals hard core. It is important to realize
that the speed of sound in the gas bubble can become relativistic at
this stage~\cite{Companion}.

We feel that theoretical computations along these lines have now been
pushed as far as is meaningful given the current experimental
situation. We have now constrained Casimir-like mechanisms for
sonoluminescence into a relatively small region of parameter space,
and it is experiment, rather than theory, that is likely to lead to
further advances. We argue, both here and elsewhere, that Casimir-like
mechanisms are viable, that they make both qualitative~\cite{2gamma}
and quantitative predictions, and that they are now sufficiently well
defined to be experimentally falsifiable. (Possible extensions of the
model will require much more detailed condensed matter information,
such as experimental data regarding the actual refractive index inside
the bubble as a function of time, space, and frequency.)

In conclusion, the present calculation (limited though it may be)
represents an important advance: There now can be no doubt that changes
in the refractive index of the gas inside the bubble lead to production
of real photons---the controversial issues now move to quantitative ones
of precise fitting of the observed experimental data.  We are hopeful
that more detailed models and data fitting will provide better
explanations of the details of the SL effect. 

\section*{Acknowledgments}

This research was supported by the Italian Ministry of Science (DWS,
SL, and FB), and by the US Department of Energy (MV). MV particularly
wishes to thank SISSA (Trieste, Italy) and Victoria University (Te
Whare Wananga o te Upoko o te Ika a Maui; Wellington, New Zealand) for
hospitality during various stages of this work.  MV wishes to thank
J.~Feinberg for perceptive questions regarding the normalization of
the static eigen-modes. SL wishes to thank Washington University for
its hospitality. DWS and SL wish to thank E.~Tosatti for useful
discussion. SL wishes to thank M.~Bertola, B.~Bassett, R.~\Schutzhold\
and G.~Plunien for comments and suggestions.
All authors wish to thank G.~Barton for his interest and encouragement.

\appendix
\section{Generalizing the inner product}

For the differential equation
\begin{equation}
\label{E:static1}
\epsilon \; \partial_{0}(\partial_{0} E)-\nabla^{2} E=0,
\end{equation}
it is a standard exercise to write down a density and flux,
\begin{equation}
\rho = \epsilon 
\left( E_1^* \; \partial_t E_2 - E_2 \; \partial_t E_1^*\right),
\end{equation}
\begin{equation}
j = E_1^* \nabla E_2 - E_2 \nabla E^*_1,
\end{equation}
and to then show that, by virtue of the differential equation
(\ref{E:static1}), these quantities satisfy a continuity equation
\begin{equation}
\partial_t \rho - \nabla \cdot j = 0.
\end{equation}
Suppose now one has two {\em solutions} of the differential equation
(\ref{E:static1}), one can then define an inner product
\begin{equation}
(E_1, E_2) = -i \; 
\epsilon \int_t \left( E^*_1 \partial_t E_2 - E_2 \partial_t E_1^*\right),
\end{equation}
where the integral is taken over a constant-time spacelike hypersurface.
By virtue of the above, this inner product is {\em independent} of
the time $t$ at which it is evaluated.

Now what happens if the dielectric is allowed to depend on both
space and time? First the differential equation of interest is
generalized to
\begin{equation}
\label{E:change1}
\partial_{0}(\epsilon(x,t) \partial_{0} E)-\nabla^{2} E=0.
\end{equation}
Second, the density and flux become,
\begin{equation}
\rho = \epsilon(r,t) 
\left( E_1^* \partial_t E_2 - E_2 \partial_t E_1^*\right),
\end{equation}
\begin{equation}
j = E_1^* \nabla E_2 - E_2 \nabla E_1^*.
\end{equation}
By virtue of the differential equation (\ref{E:change1}), 
\begin{eqnarray}
\partial_t \rho &\equiv& 
E_1^* \; \partial_t (\epsilon(x,t) \partial_t E_2) 
- E_2 \; \partial_t (\epsilon(x,t) \partial_t E_1^*)
\\
&=&
E_1^* \nabla^2 E_2 - E_2 \nabla^2 E_1^*
\\
&=&
\nabla \cdot ( E_1^* \nabla E_2 - E_2 \nabla E_1^*)
\\
&=& \nabla \cdot j.
\end{eqnarray}
Which implies that these generalized quantities satisfy a continuity
equation
\begin{equation}
\partial_t \rho - \nabla \cdot j = 0.
\end{equation}
This implies that the generalized inner-product [for two solutions
$E_1$ and $E_2$ of the equation (\ref{E:change1}) for time-dependent
and space-dependent dielectric constants] must be
\begin{equation}
(E_1, E_2) = -{i\over2} 
\int_t \epsilon(x,t) 
\left( E^*_1 \partial_t E_2 - E_2 \partial_t E_1^*\right).
\end{equation}
By the continuity equation this inner product is independent of
the time $t$ at which the integral is evaluated [provided of course,
that $E_1$ and $E_2$ both satisfy (\ref{E:change1})]. This construction
can be made completely relativistic. Define a four-vector $J^\mu$
by
\begin{equation}
J^\mu \equiv ( \rho; j^i).
\end{equation}
Then for any edgeless achronal spacelike hypersurface $\Sigma$ there is
a conserved inner product
\begin{equation}
(E_1, E_2) = -i \; 
\int_\Sigma \epsilon(x,t) \; J^\mu \; d\Sigma_\mu.
\end{equation}
%

\section{Some Bessel function identities}

The inversion formula for Hankel Integral
transforms is~\cite{Bateman,Jackson}
\begin{equation}
\label{E:hankel2}
\int_{0}^{\infty}   r \; dr \;
J_{\nu}(\kappa_1 r) \; J_{\nu}(\kappa_2 r) =
{\delta(\kappa_1 - \kappa_2) \over\sqrt{\kappa_1\;\kappa_2}}=
2 \; {\delta(\kappa_1-\kappa_2)\over(\kappa_1+\kappa_2)} =
2 \;  \delta(\kappa_1^2 - \kappa_2^2),
\end{equation}
this result being valid for $Re(\nu) > - \half$, and $\lambda_{(1,2)} > 0$.
On the other hand, another well-known standard result is
\begin{equation}
\label{E:hankel3a}
\int_{0}^{R}   r dr 
J_{\nu}(\kappa_1 r)J_{\nu}(\kappa_2 r) =
{ R 
\left[
\kappa_1 J_{\nu}'(\kappa_1 R)J_{\nu}(\kappa_2 R) - 
\kappa_2 J_{\nu}(\kappa_1 R) J_{\nu}'(\kappa_2 R) 
\right]
\over
\kappa_2^2 - \kappa_1^2
}
\end{equation}
(See, for instance (24.88) on page 142 of Spiegel~\cite{Spiegel}.
Cf.~also \cite{Abramowitz}, formula 11.3.29 on page 484.  This is
of course a special case of the more general pseudo-Wronskian
analysis presented in the text.) This is enough to imply
\begin{equation}
\label{E:hankel3b}
\lim_{R\to\infty} { R 
\left[ 
\kappa_1 J_{\nu}'(\kappa_1 R) J_{\nu}(\kappa_2 R) - 
\kappa_2 J_{\nu}(\kappa_1 R)  J_{\nu}'(\kappa_2 R) 
\right]
\over
\kappa_2^2 - \kappa_1^2
} = 
{\delta(\kappa_1 - \kappa_2) \over\sqrt{\kappa_1\;\kappa_2}}
\end{equation}
By using the asymptotic forms of the Bessel functions (see, for
example, equations (24.103)--(24.104) of Spiegel~\cite{Spiegel}) this
is equivalent to the two spectral identities
\begin{equation}
\label{E:hankel4}
\lim_{R\to\infty} { \sin(k R) \over k } = \pi \; \delta(k),
\end{equation}
and
\begin{equation}
\label{E:hankel5}
\lim_{R\to\infty} { \cos(k R) \over k } = 0.
\end{equation}
These spectral identities, together with the known asymptotic forms
of the Bessel functions, then let us generalize (\ref{E:hankel3b})
above to obtain both
\begin{equation}
\label{E:hankel6}
\lim_{R\to\infty} { R 
\left[
\kappa_1 N_{\nu}'(\kappa_1 R) N_{\nu}(\kappa_2 R) - 
\kappa_2 N_{\nu}(\kappa_1 R)  N_{\nu}'(\kappa_2 R) 
\right]
\over
\kappa_2^2 - \kappa_1^2
} = 
{\delta(\kappa_1 - \kappa_2) \over\sqrt{\kappa_1\;\kappa_2}},
\end{equation}
and  
\begin{equation}
\label{E:hankel7}
\lim_{R\to\infty} { R 
\left[
\kappa_1 J_{\nu}'(\kappa_1 R) N_{\nu}(\kappa_2 R) - 
\kappa_2 J_{\nu}(\kappa_1 R)  N_{\nu}'(\kappa_2 R) 
\right]
\over
\kappa_2^2 - \kappa_1^2
} = 0.
\end{equation}
These are the key equations needed to complete the argument leading
to the correct normalization of the {\em static} eigenmodes. [See
equation (\ref{E:normalization}).]


\end{document}